\documentclass[twocolumn,numberedappendix]{aastex6}

\usepackage{color}

\usepackage{mathtools}

\usepackage{bm}

\usepackage{float}

\usepackage{verbatim}

\newcommand{\evm}{{(-)}}
\newcommand{\evz}{{(\circ)}}
\newcommand{\evp}{{(+)}}
\newcommand{\enu}{{(\nu)}}

\newcommand{\msolar}{\mathrm{M}_\odot}

\newcommand{\boxlib}{\texttt{BoxLib}}
\newcommand{\castro}{\texttt{CASTRO}}
\newcommand{\wdmerger}{\texttt{wdmerger}}
\newcommand{\python}{\texttt{Python}}
\newcommand{\matplotlib}{\texttt{matplotlib}}
\newcommand{\yt}{\texttt{yt}}

\begin{document}

%==========================================================================
% Title
%==========================================================================
\title{White Dwarf Mergers on Adaptive Meshes\\ I. Methodology and Code Verification}

\shorttitle{WD Mergers I. Methodology}
\shortauthors{Katz et al. (2016)}

\author{Max P. Katz\altaffilmark{1}}
\author{Michael Zingale\altaffilmark{1}}
\author{Alan C. Calder\altaffilmark{1,2}}
\author{F. Douglas Swesty\altaffilmark{1}}
\author{Ann S. Almgren\altaffilmark{3}}
\author{Weiqun Zhang\altaffilmark{3}}

\altaffiltext{1}
{
  Department of Physics and Astronomy,
  Stony Brook University, Stony Brook, NY, 11794-3800, USA
}

\altaffiltext{2}
{
  Institute for Advanced Computational Sciences,
  Stony Brook University, Stony Brook, NY, 11794-5250, USA
}

\altaffiltext{3}
{
  Center for Computational Sciences and Engineering,
  Lawrence Berkeley National Laboratory, Berkeley, CA 94720
}

%==========================================================================
% Abstract
%==========================================================================
\begin{abstract}

The Type Ia supernova progenitor problem is one of the most perplexing and
exciting problems in astrophysics, requiring detailed numerical modeling to
complement observations of these explosions. One possible progenitor that has
merited recent theoretical attention is the white dwarf merger scenario, which
has the potential to naturally explain many of the observed characteristics of
Type Ia supernovae. To date there have been relatively few self-consistent simulations
of merging white dwarf systems using mesh-based hydrodynamics. This is the first
paper in a series describing simulations of these systems using a hydrodynamics
code with adaptive mesh refinement. In this paper we describe our numerical
methodology and discuss our implementation in the compressible hydrodynamics
code CASTRO, which solves the Euler equations, and the Poisson equation for
self-gravity, and couples the gravitational and rotation forces to the hydrodynamics.
Standard techniques for coupling gravitation and rotation forces to the hydrodynamics
do not adequately conserve the total energy of the system for our problem, but
recent advances in the literature allow progress and we discuss our implementation
here. We present a set of test problems demonstrating the extent to which our
software sufficiently models a system where large amounts of mass are advected
on the computational domain over long timescales. Future papers in this series
will describe our treatment of the initial conditions of these systems and will
examine the early phases of the merger to determine its viability for triggering
a thermonuclear detonation.

\end{abstract}


%==========================================================================
% Introduction
%==========================================================================
\section{Introduction}

Type Ia supernovae (SNe Ia) are among the most exciting
events to study in astrophysics. These bright, brief pulses of light
in the distant universe have led to a number of important discoveries
in recent years, including the discovery of the accelerated expansion
of the universe \citep{perlmutter1999,riess1998}. Their origin, though,
is shrouded in mystery. It has long been expected that these
events arise from the thermonuclear explosions of white dwarfs
\citep{hoyle-fowler:1960}, but the cause of these explosions is
uncertain. In particular, it is not clear what process causes the
temperatures in these white dwarfs (WDs) to become hot enough for explosive
burning of their constituent nuclei. The model favored initially by the
community was the single-degenerate (SD) model
\citep{whelan-iben:1973}. Accretion of material from a companion star
such as a red giant would cause the star to approach the Chandrasekhar
mass, and in doing so the temperature and density in the center would
become sufficient for thermonuclear fusion to proceed. In
recent years the focus has shifted to a number of alternative progenitor models. A
leading candidate for explaining at least some of these explosions is
the double-degenerate (DD) model, in which two white dwarfs merge and
the merged object reaches the conditions necessary for a thermonuclear
ignition \citep{ibentutukov:1984,webbink:1984}. Another is the double
detonation scenario, where accretion of material onto a
sub-Chandrasekhar mass white dwarf leads to a detonation inside the
accreted envelope, sending a compressional wave into the
core of the star that triggers a secondary detonation. A recent
review of the progenitor models can be found in
\citet{hillebrandt:2013}.

There are several observational reasons why double-degenerate systems
are a promising progenitor model for at least a substantial fraction
of normal SNe Ia. No conclusive evidence exists for a surviving
companion star of a SN Ia; this is naturally explained by the DD model
because both WDs are likely to be destroyed in the merger
process. Similarly, pre-explosion images of the SN Ia systems have
never clearly turned up a companion star, and in some cases a large
fraction of the parameter space for the nature of the companion star
is excluded. Additionally, not enough progenitor systems are seen for
the SD case to match the observed local SN Ia rate, whereas the number
of white dwarf binaries may be sufficient to account for this
rate. Finally, the DD model can naturally explain the fact that many
SNe Ia are observed to occur at very long delay times after the stars
were formed, since the progenitor systems only become active once both
stars have evolved off the main sequence. A thorough review of the
observational evidence about SNe Ia and further discussion of these
ideas can be found in \cite{maoz:2014}.

The first attempts to model the results of the merger process came in the
1980s. \cite{nomotoiben:1985} demonstrated that off-center carbon
ignition would occur in the more massive white dwarf as it accreted
mass near the Eddington rate from the less massive white dwarf
overflowing its Roche lobe. \cite{saionomoto:1985} tracked the
evolution of the flame and found that it propagated quiescently into
the center, converting the carbon-oxygen white dwarf into an
oxygen-neon-magnesium white dwarf. This would then be followed by
collapse into a neutron star---a result with significantly different
observational properties compared to a SN Ia. This scenario, termed
accretion-induced collapse, would be avoided only if the accretion
rate were well below the Eddington rate (see, e.g., \cite{fryer:1999}
for a discussion of the possible implications of the accretion-induced 
collapse scenario). \cite{tutukov-yungelson:1979}
observed that the collapse could be avoided if the mass loss from the secondary
was higher than the Eddington rate and thus the accreted material
formed an accretion disc, which might rain down on the primary more
slowly. The main finding was that double degenerate systems would not
obviously lead to Type Ia supernovae.

Three-dimensional simulations of merging double degenerate systems were 
first performed by \citet{benz:1990}, who used the smoothed particle
hydrodynamics (SPH) method to simulate the merger process. This was 
followed later by a number of authors 
\citep{rasio-shapiro:1995,segretain:1997,guerrero:2004,yoon:2007,loren-aguilar:2009,raskin:2012}.
The main finding of these early 3D SPH simulations was that if the 
lower-mass star (generally called the ``secondary'') was
close enough to the more massive star (the ``primary'') to begin mass
transfer on a dynamical time scale, the secondary completely disrupted
and formed a hot envelope around the primary, with a
centrifugally-supported accretion disk surrounding the core and
envelope. Carbon fusion might commence in the disk, but not at a 
high enough rate to generate a nuclear detonation. \cite{mochkovitch-livio:1990} 
and \cite{livio:2000}  also observed that turbulent viscosity in this disk 
would be sufficiently large for angular momentum to be removed from the 
disk at a rate high enough to generate the troublesome accretion 
timescales discussed by \cite{tutukov-yungelson:1979} and mentioned above. Based on this
evidence, the review of \cite{hillebrandtniemeyer2000} argued that the
model was only viable if the accretion-induced collapse problem could
be avoided. Later work by \cite{shen:2012} and \cite{schwab:2012} used
a more detailed treatment of the viscous transport in the outer
regions of the remnant and found that viscous dissipation in the centrifugally
supported envelope would substantially heat up the envelope on a  
viscous timescale, but their simulations still led to off-center carbon
burning. \cite{vankerkwijk:2010} argued that equal-mass mergers would
lead to the conditions necessary for carbon detonation in the center
of the merged object, but \cite{shen:2012} also questioned this for
reasons related to how viscous transport would convert rotational
motion into pressure support. \cite{zhu:2013} followed this with an
expanded parameter space study and argued that many of their
carbon-oxygen systems had the potential to detonate. The study of the
long-term evolution of the remnants is thus still an open subject of
research.

A recent shift in perspective on this problem started around 2010.
\cite{pakmor:2010} used the SPH method to study the merger of 
equal-mass ($0.9\ \msolar$) carbon-oxygen white dwarfs and found 
that a hotspot was generated near the surface of the primary 
white dwarf. They argued that this region had a temperature 
and density sufficient to trigger a thermonuclear
detonation. They inserted a detonation which propagated throughout 
the system. They found that the result would observationally 
appear as a subluminous Type Ia supernova. This was the first time 
a DD simulation successfully reproduced at least some characteristics of a SN
Ia. \cite{pakmor:2011} tried a few different mass combinations and
found empirically that this would hold as long as the secondary was at
least 80\% as massive as the primary. These events, where the merger
process resulted in the detonation of the system during the merger
coalescence---avoiding the much longer time-scale evolution---were
termed ``violent'' mergers.

Around the same time, however, \cite{guillochon:2010} and
\cite{dan:2011} pointed out that the previously mentioned simulations 
generally shared a significant drawback, which was that their initial conditions
were not carefully constructed. \cite{motl:2002}, \cite{dsouza:2006},
and \cite{motl:2007} (the first three-dimensional mesh-based
simulations of mass transfer in binary white dwarf systems) pioneered
the study of the long-term dynamical evolution of binary
white dwarf systems after constructing equilibrium initial
conditions. Earlier work placed the stars too close together 
and ignored the effects of tidal forces that change the shape of the 
secondary, leading to the merger
happening artificially too quickly \citep{fryer:2008}. When the initial conditions are
constructed in equilibrium, the system can be stable for tens of
orbital periods, substantially changing the character of the mass
transfer phase. One limitation of this series of studies is
that the authors used a polytropic equation of state and thus could
not consider nuclear reactions. \cite{guillochon:2010} and
\cite{dan:2011} improved on this using a realistic equation of state,
a nuclear reaction network, and a similar approach to the equilibrium
initial conditions, and found substantial agreement with the idea that
mass transfer occurs in a stable manner over tens of orbital
periods. They also found that, assuming the material accreted onto the
surface of the primary was primarily helium, explosive surface
detonations would occur as a result of accretion stream instabilities
during the mass transfer phase prior to the full merger. This could
trigger a double-detonation explosion and thus perhaps a SN Ia.

The latest violent merger developments have resulted in some possible areas of convergence.
\cite{pakmor:2012} performed a merger scenario
with a $1.1\ \msolar$ and $0.9\ \msolar$ setup, with better treatment
of the initial conditions, and indeed found that the merger process
happened over more than ten orbits. Nevertheless, they still determined
that a carbon-oxygen detonation would occur, in line with their
earlier results. \cite{moll:2014} and \cite{kashyap:2015} were also 
able to find a detonation in similarly massive systems. Notably,
the detonation occurred self-consistently and did not need to be  
intentionally triggered using an external source term.
\cite{dan:2012} and \cite{dan:2014} performed a large sweep 
of the parameter space for merger pairs and
found that pure carbon-oxygen systems would generally not lead to
detonations (and thus be violent mergers) except for the most massive
systems. They did find that for systems with WDs containing helium, many
would detonate and potentially lead to SNe Ia, either through the
aforementioned instabilities in the accretion stream, or during the
contact phase, similar to the violent carbon-oxygen WD
mergers. \cite{sato:2015} also examined the parameter space and
came to a similar conclusion for massive carbon-oxygen WD systems
(and also looked at the possibility of detonations after the
coalescence had completed), while \cite{tanikawa:2015} discussed
the plausibility of helium detonations in the massive binary case.
\cite{pakmor:2013} added a thin helium shell on their primary
white dwarf, and found that this robustly led to a detonation of the
white dwarf. For now there is preliminary support for the hypothesis
that systems with helium shells (or helium WDs), and very massive carbon-oxygen binaries,
could robustly lead to events resembling SNe Ia.

Given the considerable research into the double degenerate problem 
described above, why is another approach using a different simulation
code warranted? First and foremost, reproducibility of the results
across simulation codes and algorithms is important for gauging
confidence in this result. Most of the existing results that study 
the viability of double degenerate systems as progenitors for
Type Ia supernovae (that is, including a realistic 
equation of state and nuclear reactions) have
used the SPH method. SPH codes have a number of features which do aid
them in the study of these systems, such as conservation of
angular momentum to machine precision when there are no source terms
such as gravity (and conservation proportional to the level of
tolerance of error in the gravity solver when gravity is used).
A drawback relates to the fact that whether a prompt detonation
in a merger happens depends in detail on the nature of the
gas at the interface between the two stars, which is at much lower
density than the rest of the stellar material. The SPH codes for these
simulations generally all use
uniform mass particles, so their effective resolution is
\textit{lowest} at the stellar surface. In contrast, a code
with adaptive mesh refinement can zoom in on the regions where
hotspots will develop, while also maintaining high enough resolution
in the high-density regions to adequately capture the large-scale mass
transfer dynamics. There are also outstanding questions of
convergence in SPH (e.g.\ \citealt{zhu-SPH:2014}) and whether the method
correctly captures fluid instabilities. This is an important question
for white dwarf mergers because of the likely importance small-scale
instabilities will have on the evolution of the low-density gas at the
primary's surface. The pioneering work of \cite{agertz:2007} compared
grid and SPH codes and found some important differences. Most relevant
for this discussion is that the SPH codes could not adequately handle
mixing from the Kelvin-Helmholtz instability in the test they
propose. As pointed out by \cite{price:2008}, this is not a result of
SPH being inherently unable to model this instability, but instead it
is attributed to the fact that the standard SPH evolution equations do
not have a mechanism for capturing discontinuities in internal
energy. \citeauthor{price:2008} showed that the addition of an
artificial thermal conductivity can dramatically improve the ability
of the SPH codes to exhibit this instability. There have since been a
number of other papers discussing this issue, but to our knowledge
none of these improvements have yet been incorporated into an SPH
model of a WD merger. Another reason for caution is that other than the
most recent results of \cite{kashyap:2015}, no white dwarf merger simulation has self-consistently
resulted in a thermonuclear detonation. Reproducibility of the detonation 
through numerical simulation is critical for building 
confidence in this progenitor model.

This paper is the first in a series designed to address these
outstanding theoretical issues for white dwarf mergers. This work 
discusses the verification of our hydrodynamics code for simulating
these events. Later efforts will look at the initial conditions of the
system, the robustness with which a hotspot is found from which a
detonation could occur, and the importance of the initial white dwarf
models, which should be more sophisticated than simple carbon-oxygen
mixtures and in principle should use results from modern stellar
evolution calculations. \autoref{sec:Numerical Methodology}
describes our code and why it can provide useful results compared to
other methodologies used for this problem. 
\autoref{sec:implementation} describes the method we use for setting up a
binary white dwarf simulation. \autoref{sec:Tests} discusses a few
test problems that we use to verify that our code accurately
solves the equations of fluid dynamics. \autoref{sec:Performance}
demonstrates that the software scales well for supercomputer
applications. Finally, \autoref{sec:Conclusions and Discussion}
recaps what we have shown and highlights some of the future work we
plan to do.

%==========================================================================
% Numerical Methodology
%==========================================================================
\section{Numerical Methodology}\label{sec:Numerical Methodology}

To study the white dwarf merger problem, we use the mesh-based
hydrodynamics code \castro\footnote{\castro\ can be obtained at \url{https://github.com/BoxLib-Codes/Castro}.} \citep{castro}.
\castro\ solves the Euler
equations, along with the inclusion of optional modules for gravity,
nuclear reactions and thermodynamics. \castro\ is based on the \boxlib
\footnote{\boxlib\ can be obtained at \url{https://github.com/BoxLib-Codes/BoxLib}.}
adaptive-mesh refinement (AMR) framework \citep{rendleman:2000}, which
represents fluid data on a hierarchical mesh where regions of interest have higher
spatial resolution. \castro\ is highly parallel and is designed for
large-scale use on modern supercomputers; see 
\autoref{sec:Performance} for information on how \castro\ performs for our
problem. The next few subsections describe our approach to each of the
physics components used in this work. We direct the reader to the
original code paper for a full description of \castro's approach to
solving the equations of hydrodynamics. In this work, we report mainly
on the changes we have made to the code since its original release,
for the purpose of approaching this problem.

\subsection{Hydrodynamics}\label{sec:Hydrodynamics}

The Euler equations for hydrodynamics (in the absence of source terms) in conservative form are: 
\begin{align}
  \frac{\partial \rho}{\partial t} &= -\bm{\nabla} \cdot (\rho \mathbf{u}) \label{eq:euler_density}\\
  \frac{\partial \rho \mathbf{u}}{\partial t} &= -\bm{\nabla} \cdot (\rho \mathbf{u}\mathbf{u}) - \bm{\nabla}p \label{eq:euler_momentum}\\
  \frac{\partial \rho E}{\partial t} &= -\bm{\nabla}\cdot(\rho\mathbf{u}E + p\mathbf{u}). \label{eq:euler_energy}
\end{align}
Here $\rho$ is the mass density, $\mathbf{u} = (u, v, w)$ is the fluid velocity
vector, $p$ is the pressure, and $E = \mathbf{u}^2 / 2 + e$ is the
total specific energy, where $e$ is the internal (thermal) specific
energy (energy per unit mass).

We use the unsplit piecewise-parabolic method (PPM) solver in \castro\
to advance the hydrodynamics system in time \citep{ppmunsplit}.  A
number of changes were made to the solver, which are detailed in \autoref{app:hydro}.
These changes bring the algorithm more in line with that of
\cite{ppm}. \castro\ as originally released featured a slightly modified
version of the higher resolution limiters of
\cite{colella-sekora:2008}, which can be used in the code by setting 
\texttt{castro.ppm\_type = 2} in the inputs file (the inputs file is
a set of code parameters accessed at runtime to determine the algorithms
used in the simulation). The advantage of this limiter is that
it preserves physical extrema rather than clipping them off as in the
original approach of \cite{ppm}. Despite the advantages of this limiter 
we have found it to be unsatisfactory for our problem. There are many regions in our
problem with large density gradients (such as the interface between
the star's atmosphere and the ambient gas outside of it) and in these
regions the algorithm can yield negative densities. This often results
from the limiters interpreting these gradients as being true
minima. As a result, we use the original limiter, which is strictly
monotonicity preserving in the parabolic profiles it generates; this
is activated with \texttt{castro.ppm\_type = 1} in the inputs file.

A related issue that required a code improvement is that in cases of
large density gradients such as the edge of a star, it is possible to
generate negative densities in zones even with the more strongly
limited PPM. This can occur if a region of large density is moving
away from an ambient zone at relatively large speeds; then the net
density flux in the ambient zones can be large enough to unphysically
drag the density below zero. In practice, this occurs at the
trailing edge of a star that is moving across a grid. In such a
situation, there are two main approaches one could take: either
explicitly introduce a positivity-guaranteeing diffusive flux, or
reset the properties of the affected zone. We choose the latter
approach. Even though it is non-conservative, it preserves a
characteristic we value, which is to keep the edge of the stars
relatively sharp, as they physically should be. Since the mass of the
affected zones is typically already fairly low, this should not
seriously affect the dynamics or the energy conservation properties of our
simulation. Our strategy for a reset is as follows: when the density of 
a zone is below a pre-determined density floor (which is typically 
$10^{-5}\ \text{g cm}^{-3}$ for our stellar simulations), we look
at all adjacent zones and find the zone with the highest density.
If it is above the density floor, then we set the field values 
(density, momentum, energy, and temperature) of the
reset zone to be equal to the field values of this 
adjacent zone. If no adjacent zone reaches the density floor, then
the zone is set to the density floor, and given a temperature equal 
to the temperature floor for our simulations (which is typically 
$10^{5}\ \text{K}$ for our stellar simulations). We then recompute 
the thermodynamics to be consistent with these values. The 
velocity of the zone is set to zero. This latter approach only
occurs in very rare situations, and is there as a last resort.

\castro's approach to adaptive mesh refinement, based on its underlying
\boxlib\ framework, is to refine zones based on certain user-specified
criteria that tag regions of interest for higher spatial
resolution. Data is represented on one of a number of AMR levels,
where each level corresponds to a set of zones at the same resolution,
which covers a subset of the domain covered by the level immediately
below it. We typically call the level 0 grid the \textit{coarse} grid,
which has the lowest spatial resolution. Each finer, higher-level grid
has a higher resolution than the grid below it by some integer factor
$N$, which is restricted to be $N = 2\ \text{or}\ 4$ in \castro. The
zones are strictly contained within the rectangular extent of the
underlying coarser zones (at present, in 3D the code is restricted to representing
only Cartesian geometries with uniform spacing in each dimension). For the time
evolution of the AMR system we use subcycling, where each AMR level is
advanced at a different timestep and a correction step is applied at
the end to synchronize the various levels. The number of
subcycled timesteps is equal to the jump in refinement between levels,
so for example on a grid with three levels and two jumps of four in
refinement, the level 2 zones have 16 times higher spatial
resolution than the coarse grid and there are 16 level 2 timesteps
per level 0 timestep.

The boundary conditions on the hyperbolic system are simply
zero-gradient zones that allow material to flow directly out of the
domain. Using AMR, we make the coarse grid large enough that the
boundaries are relatively far from the region of interest. This
ensures that any boundary effects do not pollute the inner region
where the stars will eventually make contact.  We further make the
restriction that refined grids cannot reach the domain boundary.

\subsection{Microphysics}

The equation of state (EOS) for our simulations is the Helmholtz EOS
\citep{timmes-swesty:2000}. This models an electron-positron gas of
arbitrary relativity and degeneracy over a wide range of temperatures
and densities. Thermodynamic quantities are calculated as derivatives
of the Helmholtz free energy, and the values are interpolated from a
table. The natural variables of the Helmholtz free energy are
temperature and density, and calling the EOS is simplest in this
form. In hydrodynamics we often have the density and
internal energy as independent variables, and we want to obtain the
temperature, pressure, and other quantities. To do this, we employ a
Newton-Raphson iteration over the temperature (given some sufficient
starting guess) until we find the temperature that corresponds to the
desired internal energy. Sometimes this process fails to converge and
the iterative value approaches zero. In these cases we employ a
``floor'' that limits how low the temperature can go (typically 
$10^5$ K). There is a choice here how to proceed: we can either
assign this floor value to the temperature and let that zone be
thermodynamically inconsistent (the original behavior in \castro), or
we can adjust the internal energy to be thermodynamically consistent
with the temperature, at the cost of violating energy conservation. We
have found in some test problems of strong one-dimensional shocks that reach 
the temperature floor that the latter yields more accurate results. 
However, allowing the equation of state call to update the 
internal energy can actually result in significant changes to the 
total energy of the system over long periods of time, 
due not just to resets in low-density zones but also to small 
inconsistencies between the energy given to the EOS and the energy 
that is consistent with the returned temperature. These inconsistencies
are dependent on the tolerance of the Newton-Raphson iterative solve.
While this error tolerance is typically very small in an individual zone (a relative 
difference of $10^{-8}$ by default in \castro), over time and given 
a large number of zones, this can result in a significant energy 
drift. This is a serious enough problem that we opt for the energy 
conserving approach for our simulations.

\castro\ has the ability to model both nuclear reactions and radiative 
transport (in the flux-limited diffusion approximation). For all simulations 
in this paper we do not enable either, and we delay discussion of 
these modules until later papers in this series.

\subsection{Gravity}
\label{sec:gravity}

We solve the Poisson equation for self-gravity for our problem,
\begin{equation}
  \nabla^2 \Phi(\mathbf{x}) = 4\pi G\, \rho(\mathbf{x}),
\end{equation}
where $\Phi$ is the gravitational potential, $G$ is the gravitational
constant, and $\rho$ is the mass density.\footnote{In the \castro\ code, the 
right-hand side is negated and therefore $\Phi$ is positive. We use the 
sign convention that is typical for astrophysics in this paper. 
When $\Phi$ appears in the code it is negated to compensate for this.} 
The solution of this equation in \castro\ is described in \cite{castro}, and
consists of both level and composite solves, and (optionally) a final
synchronization at the end. We do not enable this final synchronization
for the merger simulations, because the grid boundaries never lie in
regions of high density, so the change in the potential due to the correction
at coarse--fine interface is always negligible.

\subsubsection{Coupling to Hydrodynamics}\label{sec:gravity_hydro_coupling}

The effect of gravity on the hydrodynamical evolution is typically
incorporated by the use of a source term for the momentum and energy
equations. In a finite volume methodology, the momentum source term 
often appears in integral form as
\begin{equation}
  \left.\frac{\partial (\rho \mathbf{u})}{\partial t}\right|_{\text{grav}} = \frac{1}{\Delta V} \int \rho \mathbf{g}\, dV
\end{equation}
and for the energy source term it is
\begin{equation}
  \left.\frac{\partial (\rho E)}{\partial t}\right|_{\text{grav}} = \frac{1}{\Delta V} \int \rho \mathbf{u}\cdot\mathbf{g}\, dV \label{eq:cell_center_gravity_source}.
\end{equation}
Here $\Delta V$ is the cell's volume.
In most hydrodynamics codes these are discretized as $\rho\,
\mathbf{g}$ and $\rho\, \mathbf{u}\,\cdot\mathbf{g}$, respectively, 
where $\rho$, $\mathbf{u}$, and $\mathbf{g}$ 
are evaluated at the zone center. 

There are two ways that these source terms enter the system evolution. 
First, during the hydrodynamics update, we alter the edge states that enter
into the determination of the fluxes. (This only applies for the momentum source term;
the gravitational force does not directly do work on the internal energy, which is used 
to infer the pressure.) To second order in space and time, 
this can be done using the cell-centered
source term evaluated at time-level $n$. We choose a more accurate approach, 
which is also second order, of characteristic tracing
under the source term; the details of this are described in 
\autoref{app:hydro}. Second, after the hydrodynamics step, we add the time-centered source terms
to the state. First we describe how we do this for the momentum,
and then we describe our approach for the energy. This discussion is somewhat detailed.
We believe that the attention is necessary because of the importance of accuracy
in the gravitational source terms for our problem. The stability of the white dwarf binary
system is dependent in large part upon accurate coupling of the hydrodynamics and gravity;
an error in this approach could lead to, for example, a spurious mass transfer episode
that might lead us to very different conclusions about the long term stability of such a system.
Such considerations are generally unimportant for spherically-symmetric single star calculations,
but are of the utmost importance in a simulation where the global gravitational field can change 
quite significantly over the course of the simulation.

In a system with self-gravity, total momentum is conserved if the spatial domain
includes all of the mass of the system. This must be the 
case because each mass element exerts an equal and opposite gravitational force 
on every other mass element. However, the standard approach does not necessarily
guarantee that momentum is conserved numerically. We cannot represent a vacuum state 
in our code, so there is a small but non-zero density on the edge of the grid. 
This allows momentum to leak out of the domain even if the gravitational source term 
is written in an explicitly conservative manner. To see this, one can use the Poisson equation to write the 
density in terms of the potential and then consider its spatial discretization. For simplicity,
we consider one spatial dimension and a uniform discretization. Analogous results 
may be readily obtained for the non-uniform case.
\begin{align}
  -\rho_{i}  \frac{d\Phi_{i}}{dx} &= -\frac{1}{4\pi G} \frac{d^2\Phi_i}{dx^2} \frac{d \Phi_i}{dx} \notag \\
  &= -\frac{1}{4\pi G} \left[\frac{\Phi_{i-1} - 2 \Phi_{i} + \Phi_{i+1}}{\Delta x^2}\right] \left[ \frac{\Phi_{i+1} - \Phi_{i-1}}{2\Delta x} \right] \notag \\
  &= -\frac{1}{8\pi G \Delta x^3} \left[ \Phi_{i+1}^2 - \Phi_{i-1}^2 - 2\Phi_i\left(\Phi_{i+1} - \Phi_{i-1}\right) \right] \label{eq:momentum_discretization}
\end{align}
It is easy to verify that adding the source terms for the current zone and the two zones 
to the left and right results in complete cancellation of the source terms.
The catch is that if the potential if non-zero outside of the domain, then there will be
momentum lost or gained from the grid, which will be encapsulated in the ghost cells
just outside the domain. In addition, when we replace the Laplacian above by the full
three-dimensional stencil including the $y$ and $z$ derivatives, depending on the
discretization these may not be cancelled at all. This latter problem can be resolved by
writing the momentum update in an explicitly conservative way.

\citet[Chapter 4]{shu:1992} observes that it is possible to describe the source term 
for the momentum equation by taking the divergence of a gravitational stress tensor,
\begin{equation}
  G_{ij} = -\frac{1}{4\pi G}\left(g_i g_j - \frac{1}{2}|\mathbf{g}|^2\delta_{ij}\right).
\end{equation}
The momentum equations are then written explicitly in conservative form.
The flux at any zone boundary is added to one cell and
subtracted from another, so that the total momentum in the domain interior stays constant to
within numerical roundoff error. This result can be derived by analytically recasting 
\autoref{eq:momentum_discretization}. In the continuum limit, the two momentum
formulations are identical. Thus the latter has been advocated by, for example, 
\cite{jiang:2013} for the ATHENA code. A significant limitation to this approach is that in a finite discretization 
the divergence of the gravitational acceleration is no longer guaranteed to equal
the zone density. In particular, we find that the mixing of the gravitational accleration components
means that the truncation error in the gravitational field can lead to large errors
that imply a density much different than the zone's actual density. This is especially
problematic in a simulation with a low-density ambient medium, where even a small error 
in the momentum update can lead to large changes in a zone's momentum. By continuing to explicitly
use the cell density in the momentum update, we can avoid this possibility: the size of the update
will always be suitably small if the zone's density is small. Thus for our simulations
we continue to use the standard source term for the momentum.

Time centering of this source term is done in \castro\ using a predictor-corrector approach.
At the start of a coarse grid timestep, we solve the gravitational potential for the density $\rho^n$.
We then add to the momenta a prediction of the source term that is first-order accurate in time, 
$\Delta t\, \rho^n\, \mathbf{g}^n$. After the hydrodynamics update, we recalculate
the gravitational potential based on the new density, $\rho^{n+1}$, and then add 
$-(\Delta t/2) \rho^n \mathbf{g}^n + (\Delta t/2) \rho^{n+1} \mathbf{g}^{n+1}$ to the momenta.

For the energy equation, the central challenge is to write down a form of the 
discretized energy equation that explicitly conserves total energy when 
coupled to gravity. When gravity is included, the conserved total energy
over the entire domain is
\begin{equation}
  \int \rho E_{\text{tot}}\, dV = \int dV \left(\rho E + \frac{1}{2}\rho\Phi\right), \label{eq:total_energy_gravity}
\end{equation}
where $\rho E$ is the total gas energy from the pure hydrodynamics equation. 
The factor of 1/2 in the gravitational energy is necessary for simulations with
self-gravity to prevent double-counting of interactions (since in dynamical evolution
the relevant gravitational potential energy is $\rho \Phi$ and the gravitational force
is $\rho \mathbf{g}$). Historically many simulation codes with gravity have not used
a conservative formulation of the energy equation, but it is straightforward to do so.
Our approach, and the discussion that follows, is based on that of \cite{arepo}.

Conservation of total energy requires that a change in gravitational energy is compensated
for by a change in gas energy, and that energy changes due to mass transfer are explicitly and 
exactly tracked. Suppose that we have some fluid mass $\Delta M_{i+1/2} = \Delta \rho_{i+1/2} \Delta V$ leave the zone
with index $i$ and enter the zone with index $i+1$. The subscript indicates that the mass change is
occurring at the interface between the two zones, at index $i+1/2$. The work done by the gravitational
force on the gas is $\Delta (\rho E) = W = \int F dx = (\Delta M_{i+1/2}\ g_{i+1/2}) (\Delta x / 2)$,
where $g_{i+1/2}$ is the gravitational acceleration at the interface. The second term in parentheses
is just the distance from the zone center to the zone edge: once the mass leaves the zone edge, it no longer
needs to be tracked. To second order, $g_{i+1/2} = -(\Phi_{i+1} - \Phi_{i}) / \Delta x$, and also to second order the potential
at the interface is given by $\Phi_{i+1/2} = (\Phi_{i+1} + \Phi_i) / 2$, so we can equivalently view the work done
as $W = -\Delta M_{i+1/2} (\Phi_{i+1/2} - \Phi_i)$. Physically, this is just the negative of the gravitational
potential energy change as the fluid is pushed from the cell center potential to the cell edge potential,
exactly as the work-energy theorem implies. 

Now, in a hydrodynamics code, mass changes correspond to hydrodynamic fluxes. In particular,
the continuity equation tells us that the mass flux $F_\rho = \rho^{n+1/2}_{i+1/2} v^{n+1/2}_{i+1/2}$ yields
an integrated mass motion through the interface $i+1/2$ over a timestep $\Delta t$ of:
\begin{equation}
  \Delta \rho_{i+1/2} = \frac{\Delta t}{\Delta V} \left(\rho^{n+1/2}_{i+1/2} v^{n+1/2}_{i+1/2} dA\right).
\end{equation}
Note that here $v_{i+1/2}$ is the component of the velocity perpendicular to the zone face, whose
area is $dA$.

Finally, then, we write the update in a zone for the total energy that conserves $(\rho E_{\text{tot}})$ as:
\begin{equation}
  \Delta (\rho E) = -\frac{1}{2}\sum_{f} \Delta \rho_{f} (\Phi_{f+1/2} - \Phi_{f-1/2}),\label{eq:grav_energy_conservation_update}
\end{equation}
where the sum is over the cell faces with indices $f$ and the indices $f+1/2$ and $f-1/2$ refer to 
the zone centers immediately to the left and right in the direction perpendicular to the face.
As long as we record the hydrodynamical fluxes through the zone faces after coming out of the hydrodynamics step, 
this algorithm is able to conserve the total energy completely (except for any energy loss or gain through 
physical domain boundaries). In order for the method to be second-order accurate in time, 
we need to use a time-centered $\Phi$ (which can be computed by averaging the time-level $n$ and $n+1$ potentials;
we already have the latter because \castro\ re-computes the potential at the new time after the hydrodynamics step,
and we can apply this energy at the end of the timestep). Note that of course the hydrodynamical
flux is already second-order accurate in time. We observe also that in practice we will not obtain 
conservation of energy to machine precision even in the absence of open domain boundaries. The 
method itself is conservative if it is time-centered and correctly evaluates the energy change 
on cell faces. This was demonstrated empirically by \cite{jiang:2013} and is obvious in the case of a
fixed external potential; it is not as obvious in the case of the gravitational self-potential, which
changes in response to changes in the mass distribution, so we give a short proof of this in
\autoref{app:gravity}. However, in practice there is a non-zero numerical tolerance associated 
with the Poisson gravity solver (in our case, the multigrid method) that results in a non-zero error 
in the calculation of the gravitational potential. This results in a very small deviation from perfect 
conservation. It is not usually larger than the other effects which result in energy non-conservation 
for our simulations, such as resetting the state of zones that acquire a negative internal energy, and 
in principle if desired it can be made smaller by using stricter tolerance levels on the gravity solve.

In passing, we hope to clear up a spot of potential confusion, that we feel is unclear in other papers
on this subject: the factor of $1/2$ that appears in \autoref{eq:grav_energy_conservation_update} 
has nothing to do with the factor of $1/2$ that appears in the statement of conservation of total energy, 
\autoref{eq:total_energy_gravity}. The former comes simply from the fact that the energy change is 
evaluated using the mass motion through a distance of half of the zone width. The latter is needed 
to ensure that these local changes in energy are not double-counted when doing a global integral, 
since the gravitational potential is self-generated. \autoref{eq:grav_energy_conservation_update} 
applies to any conservative potential $\Phi$, and we use this to our advantage for the 
rotation forces in \autoref{sec:rotation}.

As observed by \cite{arepo}, this method is more accurate than the more common (non-conservative) approach
of evaluating the change in gas energy using the work done $(\mathbf{v} \cdot \rho \mathbf{g})$
by the gravitational force at the cell center. Analytically this form expresses the same core idea as
\autoref{eq:grav_energy_conservation_update} via the work-energy theorem, but a major flaw is that
it evaluates the energy change at the cell center when in fact the mass transfer is happening at
the cell edges. This can result in a significant leaking of energy throughout the course of the
evolution, dramatically affecting the course of the evolution. The standard approach is therefore
unacceptable in the case of a problem like white dwarf mergers, and the fix to this energy
leaking---evaluating the energy transfer at the six zone faces instead of the single zone
center---adds only a very minor cost in terms of code complexity and computational time.

Another approach to conserving total energy recently taken in the literature is to evolve an 
equation for the total energy $(\rho E_{\text{tot}})$; see \cite{jiang:2013} (see also 
\cite{arepo}, Section 5.3). That is, one can replace the gas energy equation with a total energy equation, 
and then the energy flux includes a term corresponding to the flux of gravitational potential energy. We 
avoid this approach for our problem because there are regions on the computational domain where the total 
energy is dominated by potential energy (especially the low-density regions near the edge of the white dwarfs),
and the gas energy can only be retrieved by first subtracting $-\rho \Phi/2$ from the total energy. Like 
\cite{arepo}, we find that this can result in some serious errors due to numerical discretization, yielding 
unphysical energies or temperatures. We observe also that the implementation of \cite{jiang:2013} 
has terms in the gravitational flux that are not proportional to $\rho$, and so can lead to the 
same troubles that plague the tensor-based formalism for the momentum equation, where small errors 
in the discretization of the gravitational potential can lead to very large changes in the energy of the gas.

\subsubsection{Boundary Conditions}\label{sec:gravity_boundary_conditions}

Analytical solutions to the Poisson equation customarily assume that the
potential vanishes at large distances from the region of non-zero
density. On a finite computational domain, however, it is usually not
possible to have the edges of the domain be far enough away that the
potential can be taken to be zero there. Solving the Poisson equation
therefore requires knowledge of the values of the potential on the
edges of the computational domain. In principle, the boundary values can be computed
by doing a direct sum over the mass distribution inside the domain,
where the mass in each zone is treated as a point source:
\begin{equation}
  \Phi_{{lmn}} = -\sum_{{i, j, k}} \frac{G \rho_{{ijk}}}{|\mathbf{x}_{{lmn}} - \mathbf{x}_{{ijk}}|}\, \Delta V_{{ijk}}.\label{eq:direct_sum}
\end{equation}
Here $(i, j, k)$ are the indices of cells inside the domain, and $(l,m, n)$ 
are the indices of ghost zones outside the domain where the boundary values for the potential is specified\footnote{In \castro\ we actually
specify the potential on cell edges, not on cell centers, but the idea is the same, and we use the location of
the cell edge in computing the distance to each zone in the domain.}. $\Delta V$ is the volume of the
zone. If there are $N$ zones per spatial dimension, then there are
$6 N^2$ boundary zones, and each boundary zone requires a sum over
$N^3$ zones, so the direct computation of the boundary conditions
scales as $\mathcal{O}(N^5)$.  This method is expensive enough that it is not used
for hydrodynamics simulations (though it is useful for comparison to
approximate solutions, so we have implemented it as an option in \castro).

In a typical simulation we place the boundaries of the domain far
enough away from the region containing most of the mass that some
method of approximation to this direct summation is justified. Many
approaches exist in the literature. The original release of \castro\
featured the crudest possible approximation: a monopole prescription,
where the boundary values were computed by summing up all the mass on
the domain and treating it as a point source at the domain
center. This is correct only for a spherically symmetric mass
distribution, and therefore is best suited for problems like
single-star Type Ia supernova simulations (e.g.$\ $\cite{malone:2014})
that employ self-gravity. For a problem like that of a binary star system
with significant departures from spherical symmetry, this assumption
fails to produce accurate boundary values, which we find in \autoref{sec:kepler}
results in a significant drift of the center of the mass of the system over time.

The most natural extension of the monopole prescription is to include
higher-order multipole moments. If the entire mass distribution is
enclosed, then the potential can be expanded in a series of spherical
harmonics $Y_{lm}(\theta,\phi)$ (where $\theta \in [0, \pi]$ is the usual polar angle
with respect to the $z$ axis and $\phi \in [0, 2\pi)$ is the usual azimuthal angle with
respect to the positive $x$ axis):
\begin{equation}
  \Phi(\mathbf{x}) = -\sum_{l=0}^{\infty}\sum_{m=-l}^{l} \frac{4\pi}{2l + 1} q_{lm} \frac{Y_{lm}(\theta,\phi)}{r^{l+1}}, \label{eq:spherical_harmonic_expansion}
\end{equation}
where $q_{lm}$ are the so-called multipole moments. The origin of the
coordinate system is taken to be the center of the computational
domain, and $r$ is the distance to the origin. The multipole moments
can be calculated by expanding the Green's function for the Poisson
equation as a series of spherical harmonics. After some algebraic
simplification of \autoref{eq:spherical_harmonic_expansion}, 
the potential outside of the mass distribution can be written as:
\begin{align}
  \Phi(\mathbf{x}) &= -\sum_{l=0}^{\infty} \left\{Q_l^{(0)} \frac{P_l(\text{cos}\, \theta)}{r^{l+1}} \right. \notag \\
    &+ \left. \sum_{m = 1}^{l}\left[ Q_{lm}^{(C)}\, \text{cos}(m\phi) + Q_{lm}^{(S)}\, \text{sin}(m\phi)\right] \frac{P_{l}^{m}(\text{cos}\, \theta)}{r^{l+1}} \right\}.\label{eq:multipole_potential}
\end{align}
$P_l(x)$ are the Legendre polynomials and $P_{l}^{m}(x)$ are the associated Legendre polynomials.
$Q_l^{(0)}$ and $Q_{lm}^{(C,S)}$ are variants of the multipole moments that involve integrals of
$P_l$ and $P_l^m$, respectively, over the computational domain; their definition is given in \autoref{app:multipole}.

This approach becomes computationally feasible when we cut off the
outer summation in \autoref{eq:multipole_potential} at some finite
value of $l_{\text{max}}$. If it is of sufficiently high order, we
will accurately capture the distribution of mass on the grid. In
practice we first evaluate the discretized analog of the modified
multipole moments for $0 \leq l \leq l_{\text{max}}$ and $1 \leq m
\leq l$, an operation that scales as $N^3$. We then directly compute
the value of the potential on all of the $6N^2$ boundary zones. Since
the multipole moments only need to be calculated once per Poisson
solve, the full operation scales only as $N^3$. The amount of time
required to calculate the boundary conditions is directly related
to the chosen value of $l_{\text{max}}$, so there is a trade-off
between computational expense and accuracy of the result.

\begin{figure}[h]
  \centering
  \includegraphics[scale=0.45]{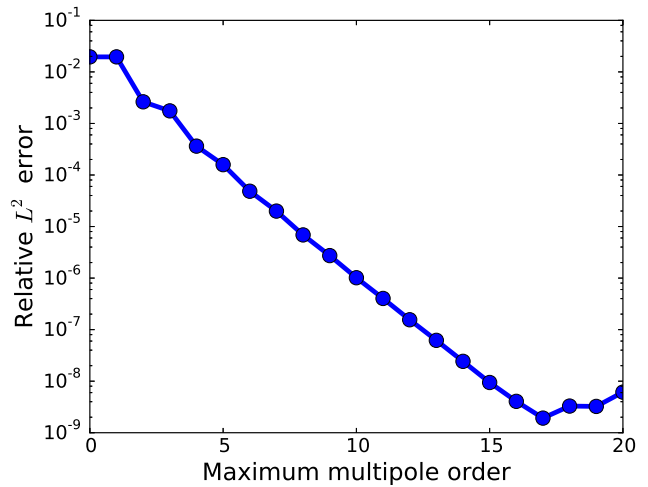}
  \caption{Error of $\Phi$ on the computational domain for a binary white dwarf simulation 
    whose boundary conditions were computed using various values of the maximum multipole order,
    relative to the exact solution determined by a brute force sum on the boundaries.
    Circles represent the error at integer values, and they have been connected by a smooth 
    line to guide the eye.\label{fig:bc_comparison}}
\end{figure}

As a demonstration of the method's accuracy, we consider the case of two 
white dwarfs of mass ratio 2/3, using the initialization procedure described below 
in \autoref{sec:implementation}. We terminated the simulation just after
initialization, so that we perform only an initial Poisson solve for this 
density distribution. We did this for values of $l_{\text{max}}$ ranging
from 0 to 20, and we also did this using the numerically exact solution 
provided by \autoref{eq:direct_sum}.  Defining the $L^2$
norm of a field $f$ as
\begin{equation}
  \| f \|_2 = \left(\sum_{i,j,k} \Delta x\, \Delta y\, \Delta z\, f_{ijk}^2\right)^{1/2},
\end{equation}
we computed the $L^2$ error of $\Phi$ on the entire domain for multipole 
boundary conditions, which we call $\Phi_l$, relative to $\Phi$ 
obtained using the exact boundary conditions:
\begin{equation}
  \text{Error}_l = \frac{\|\Phi_l - \Phi_{\text{exact}}\|_2}{\|\Phi_{\text{exact}}\|_2}.
\end{equation}
The result is shown in \autoref{fig:bc_comparison}. At $l_{\text{max}} = 6$,
the error is already well below $10^{-4}$ and we adopt this as our default 
choice for all simulations with Poisson gravity. In \autoref{sec:kepler} we 
show that there are no gains to be had by increasing the accuracy further. At 
very high orders ($l \gtrsim 18$) the approximation breaks down, as seen in \autoref{fig:bc_comparison}. 
This is a result of the ambient material on the grid. At each boundary point we 
assume that all of the mass on the grid is contained within a sphere whose radius is 
the distance from that boundary point to the center of the domain. This does not hold 
for boundary points in the centers of domain faces, because of the material in the 
domain corners. This can be fixed by using multiple mass shells at diferent radii, but the error 
is negligible in practice for the values of $l_{\text{max}}$ that we use.

\subsubsection{Convergence Testing}\label{sec:gravity_convergence_testing}

\begin{figure}[h]
  \centering
  \includegraphics[scale=0.45]{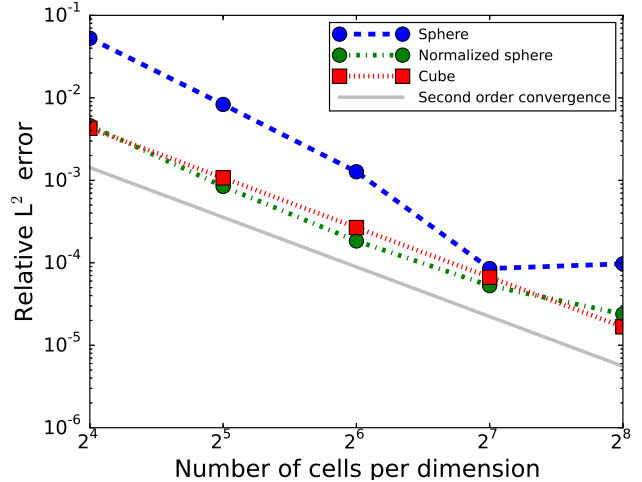}
  \caption{Comparison of the \castro\ gravitational potential to the analytical solution for: 
    a sphere of uniform density; the same sphere, but with the potential normalized using the 
    actual amount of mass on the grid instead of the mass of a perfect sphere; and, a 
    cube of uniform density. Plotted also is a notional curve whose slope represents
    perfect second order convergence.\label{fig:gravity_convergence}}
\end{figure}

Since the results of a merger simulation depend strongly on gravity,
it is important to check whether proper numerical convergence is
achieved for the Poisson solver. To do so, we created a simple test
that initializes a sphere of radius $R$ and uniform mass density $\rho$
onto our grid, and used \castro\ to calculate the gravitational
potential $\Phi$ of this setup. We ensure that $R$ is an integer
multiple of the grid spacing, and the center of the sphere is at the
origin. The problem domain for our simulations is $[-1.6\ \text{cm}, 1.6\ \text{cm}]^3$, and
we take $R = 1.0\ \text{cm}$ and $\rho = 10^3\ \text{g cm}^{-3}$. 
The zones with $r > R$ are filled with an ambient material of very low density 
($10^{-8}\ \text{g cm}^{-3}$). We run this problem at multiple 
resolutions corresponding to jumps by a factor of two. For
comparison, at each grid point we evaluate the analytical potential of
a uniform sphere, which can be easily determined using Gauss' law:
\begin{equation}
  \Phi_{\text{sphere}}(r) = -\frac{GM}{r} \times \begin{cases} (3R^2 - r^2)/(2 r^2) & r \leq R \\ 1 & r > R \end{cases},\label{eq:sphere-analytical}
\end{equation}
where $M = 4\pi R^3 / 3$ is the mass of the sphere. We measure the 
numerical error by calculating the $L^2$ norm of the error and 
normalizing it by the $L^2$ norm of the analytical solution:
\begin{equation}
  \text{Error} = \frac{\|\Phi - \Phi_{\text{sphere}}\|_2}{\|\Phi_{\text{sphere}}\|_2}.
\end{equation}
We define the order of convergence $p$ between two simulations with a jump 
in resolution of integer factor $m > 1$ as
\begin{equation}
  p = \text{log}_{m}\left(\frac{\text{Error}_{\text{low}}}{\text{Error}_{\text{high}}}\right).
\end{equation}
Here $\text{Error}_{\text{low}}$ is the $L^2$ error at the lower resolution 
and $\text{Error}_{\text{high}}$ is the $L^2$ error at the higher resolution.
We expect the error to converge at $p = 2$ given the discretization we choose. 
For all simulations in this section and for all our main science simulations,
we choose a relative error tolerance of $10^{-10}$ to be satisfied in the multigrid solve.
The results of this test are plotted in \autoref{fig:gravity_convergence}. 

We find that at low resolution convergence is actually substantially better 
than second-order. The explanation for this is that we are attempting to 
model a spherical object on a rectangular grid. This results in two sources of error.
First, at very low resolution, the object does not look very spherical due to the rectangular 
grid representation, so the potential it produces is not quite that of a sphere. 
As the resolution is increased, the distribution of the mass on the grid will change.
Second, the total amount of mass on the grid will change as the sphere fills out. 
So we are combining the true accuracy bonus from increased resolution 
with the artificial accuracy bonus from getting closer to solving the problem 
we are supposed to be solving. At high resolution this effect levels off, though, 
as the representation of the sphere is not significantly different in 
our two highest resolutions shown. For example, at $128$ zones per dimension 
the amount of mass on the grid happens to be slightly closer to the true spherical 
mass than at $256$ zones per dimension.
We can eliminate the second source of error by changing the density on  
the grid so that the total mass $M$ is actually what we intend it to be.
The resolution study for this case (the ``normalized sphere'') is also
plotted in \autoref{fig:gravity_convergence}. At low resolution we still obtain
convergence slightly better than second-order, indicating that we 
have not eliminated the geometrical problem of the mass distribution changing.

The only way to fully eliminate this effect is to use a test problem that
does not change with resolution. The obvious companion problem is a cube of
uniform density $\rho$, where now $R$ is half of the side length of
the cube. At each resolution we use the same $R$ as for the sphere,
which ensures that the cube always fills exactly the same fraction of
the domain and thus has the same mass, so the only improvement comes
from better sampling at higher resolution. The gravitational potential for this
object has been worked out analytically by \citet{waldvogel:1976} (see
also a similar result by \citet{hummer:1996}, and an earlier calculation 
by \citet{macmillan:1958}). The potential is given in
Equation 15 of that paper\footnote{The last term in that equation is missing a factor of
$1/2$, which destroys the symmetry. We have inserted this missing factor and
performed a simple coordinate transformation so that the center of
the cube is at the origin.}:
\begin{align}
  \Phi_{\text{cube}}(x,y,z) &= -G\rho\sum_{i,j,k=0}^1\left[x_i y_j\, \text{tanh}^{-1}\left(\frac{z_k}{r_{ijk}}\right)\right. \notag \\
  &+ \left. y_j z_k\, \text{tanh}^{-1}\left(\frac{x_i}{r_{ijk}}\right) + z_k x_i\, \text{tanh}^{-1}\left(\frac{y_j}{r_{ijk}}\right) \right.\notag \\
  &\left. - \frac{x_i^2}{2}\,\text{tan}^{-1}\left(\frac{y_j z_k}{x_i r_{ijk}}\right) - \frac{y_j^2}{2}\,\text{tan}^{-1}\left(\frac{z_k x_i}{y_j r_{ijk}}\right) \right. \notag \\
  &- \left. \frac{z_k^2}{2}\,\text{tan}^{-1}\left(\frac{x_i y_j}{z_k r_{ijk}}\right)\right]
\end{align}
where $x_0 = R + x$, $x_1 = R - x$, $y_0 = R + y$, 
$y_1 = R - y$, $z_0 = R + z$, $z_1 = R - z$, 
and $r_{ijk} = \sqrt{x_i^2 + y_j^2 + z_k^2}$. We note that if implemented in 
\texttt{Fortran} or \texttt{C}/\texttt{C++}, the inverse hyperbolic tangent used here is
\texttt{atanh} and the inverse tangent is \texttt{atan} (\textit{not}
\texttt{atan2}). This formula is valid both inside and outside the
cube. The normalized $L^2$ error for this problem is also shown
in \autoref{fig:gravity_convergence}, and only for this problem 
do we obtain perfect second-order scaling at all resolutions.

The main lesson here is that in a convergence study, it is important
to ensure that the physical problem does not change with
resolution. Since in the case of spherical objects on rectangular
grids the effect may be to artificially boost convergence with resolution,
in a simulation with spherical objects like stars one can envision a
scenario of being fooled into believing apparently good convergence
results that are simply a convolution of artificially high
gravitational convergence and poor convergence in the hydrodynamics. A
convergence study in this case is only fully valid if there is reason
to be confident that this effect is negligible compared to other
factors.

\subsection{Rotation}\label{sec:rotation}

For the evolution of binary systems, it is most natural to evolve the
two stars in a frame that is co-rotating at the same period as the
orbital period. Since the publication of the original code paper, \castro\ 
now has the ability to evolve systems in a rotating reference frame. 
Source terms corresponding to the Coriolis and centrifugal 
force terms are added to the momentum and energy equations. In this frame, 
the stars essentially remain stationary in their original positions due to the
centrifugal force supporting against the gravitational attraction, and
will remain this way as long as significant mass transfer does not
occur. \cite{swc:2000} demonstrated (in the context of neutron star
mergers) that conservation of angular momentum is much easier to
obtain in the rotating reference frame than in an inertial frame in
which stars advect large amounts of material around the domain. We
wish to emphasize that although it is commonly stated in the
literature that fixed-mesh codes poorly conserve angular momentum,
it is only generally true that mesh-based codes do not exactly conserve 
angular momentum when the equations are written in conservative form
for linear momentum. Indeed, \cite{motl:2002} and \cite{byerly:2014} 
have evolved binary systems using the hydrodynamics equations written 
in a form that explicitly conserves angular momentum, and it is 
straightforward to convert an existing grid-based code to solve 
the system of equations that \citeauthor{byerly:2014} present.
Additionally, the extent to which angular momentum conservation is violated in our code
is a function of the resolution. When the resolution is sufficiently high, 
excellent conservation properties can result. At reasonable resolution 
for a binary orbit our code conserves angular momentum well enough 
to keep the stars stable for a large number of orbits; however, at moderate 
resolution in an inertial frame, there is a secular loss of angular 
momentum that eventually will result in a spurious merger.
We note that as the stars begin to coalesce, the rotating reference frame
will no longer provide a good approximation to the spatial motion of
the stars and then they will begin to significantly move around the
domain. This is not necessarily problematic because the most important
feature of the rotating frame is that it helps ensure that the initial
coalescence is not the result of spurious numerical loss of angular
momentum. When significant mass transfer sets in and evolution
proceeds on a dynamical timescale, the conservation properties may be
slightly worse but angular momentum conservation is also less
important.

In a rotating reference frame with angular frequency vector $\bm{\omega}$,
the non-inertial contribution to the momentum equation is:
\begin{equation}
  \left.\frac{\partial(\rho \mathbf{u})}{\partial t}\right|_{\text{rot}} = -2\, {\bm\omega} \times (\rho\mathbf{u}) - \rho {\bm\omega} \times \left({\bm\omega} \times \mathbf{r}\right).
\end{equation}
Here $\mathbf{r}$ is the position vector with respect to the origin. Typically we choose $\bm{\omega} = (0, 0, 2\pi / T)^T$,
with the rotation axis coincident with the $z$ axis at $x = y = 0$.
$T$ is the rotation period, which is the most natural quantity to specify
for a rotating stellar system. As described in \autoref{app:hydro}, we include this source term
in the edge state prediction in a way that is analogous to the gravity source.
We evaluate all quantities at cell centers. We use the same predictor-corrector 
approach that we use for the gravity source terms to the momentum equations. A slight 
difference is that the Coriolis force for each velocity component is coupled to other velocity 
components. If the rotation is about the $z$-axis, then the discrete update to 
$u^{n+1}$ depends on the value of $v^{n+1}$, and vice versa. If we fix the value of 
the time-level $n+1$ quantities after coming out of the hydrodynamics update, there 
would be a slight inconsistency between the $x$ and $y$ components of the velocity. 

We propose a more accurate coupling that directly solves this implicit system of coupled 
equations. We denote by $(\widetilde{\rho \mathbf{u}})$ the value of the momentum after 
updating it with the centrifugal force, and the time-level $n$ Coriolis force. The remaining 
update for the time-level $n+1$ Coriolis force then appears as:
\begin{equation}
  (\rho \mathbf{u})^{n+1} = (\widetilde{\rho\mathbf{u}}) + \frac{\Delta t}{2} \left(-2\, {\bm\omega} \times (\rho\mathbf{u})^{n+1}\right)
\end{equation}
To proceed further, we assume that the rotation is about the $z$ axis with frequency $\omega$. 
Then there is no update to the $z$-momentum, and the other equations are:
\begin{align}
  (\rho u)^{n+1} &= (\widetilde{\rho u}) + \omega \Delta t (\rho v)^{n+1} \\
  (\rho v)^{n+1} &= (\widetilde{\rho v}) - \omega \Delta t (\rho u)^{n+1}
\end{align} 
We can directly solve this coupled system:
\begin{align}
  (\rho u)^{n+1} &= \frac{ (\widetilde{\rho u}) + \omega \Delta t (\widetilde{\rho v})}{1 + \omega^2 \Delta t^2} \\
  (\rho v)^{n+1} &= \frac{ (\widetilde{\rho v}) - \omega \Delta t (\widetilde{\rho u})}{1 + \omega^2 \Delta t^2}
\end{align}
We use this form of the momentum update in \castro. This improvement is small
but increases the accuracy of our rotating white dwarf systems over long time-scales.

The update to the energy equation can be determined by taking the dot product of the velocity
with the momentum source terms. The Coriolis term vanishes identically, and so
the Coriolis term does no work on the fluid. The update from the centrifugal force becomes
\begin{equation}
  \left.\frac{\partial(\rho E)}{\partial t}\right|_{\text{rot}} = \frac{1}{\Delta V}\int \rho \mathbf{u} \cdot \mathbf{f}^R\, dV,
\end{equation}
with $\mathbf{f}^R \equiv  -{\bm\omega} \times \left({\bm\omega} \times \mathbf{r}\right)$. 
This expression is identical in form to the gravity source under the interchange of $\mathbf{g}$ with $\mathbf{f}^R$.
As observed by \cite{marcello:2012}, we can similarly write down a rotational potential,
\begin{equation}
  \Phi^R = \frac{1}{2} \left| {\bm\omega} \times \mathbf{r} \right|^2.
\end{equation}
In the presence of rotation the conserved total energy becomes:
\begin{equation}
  \int dV (\rho E_{\text{tot}}) = \int dV \left( \rho E + \frac{1}{2} \rho \Phi + \rho \Phi^R \right).
\end{equation}
Given that we can write down a potential energy for the rotation field, then we can use the machinery of 
\autoref{sec:gravity_hydro_coupling}. We again continue to evolve explicitly an equation for 
the gas energy, and allow it to change in response to work done by or on the rotational potential.
\begin{align}
  \left.\Delta(\rho E)\right|_{\text{rot}} &= -\frac{1}{2}\sum_{f} \Delta \rho_{f} (\Phi^R_{f+1/2} - \Phi^R_{f-1/2})
\end{align}

We apply the rotational forces after the gravitational forces, but 
there is some freedom in the order in which to apply the gravitational and rotational terms.
This order may matter because the Coriolis force depends on the fluid velocity, and 
in the predictor-corrector approach, we use the velocities both at 
time-level $n$ and time-level $n+1$. If we update the latter with the gravitational force, 
then the Coriolis force sees a different velocity than the one obtained through the 
pure hydrodynamics step. (The energy equation does not face the same issue in our new formulation,
because the velocities used are always the time-level $n+1/2$ values coming from the Riemann solver.)
In practice, this does not matter significantly for our simulations in this work 
because the centrifugal force plays the dominant role in maintaining stability of non-contact 
binary systems, and the centrifugal force does not depend on the fluid velocity.
This issue may be worth exploring in future work in situations where the Coriolis 
term is non-negligible in determining the system evolution.

In all simulations performed in a rotating reference frame, we transform all relevant
quantities back to the inertial reference frame when reporting them in analysis routines 
and visualization (though the data is saved to plotfiles while still in the rotating frame). In particular,
for every zone we adjust the position, momentum, and energy to account for rotation.
If the position is $\mathbf{x}$ in the inertial frame and $\mathbf{x}^\prime$ in 
the rotating frame, and the rotation vector is $\bm{\omega}$, the transformation rules are:
\begin{align}  
  \mathbf{x}(t) &= \mathbf{R}\mathbf{x}^\prime(t) \\
  \mathbf{v}(t) &= \mathbf{v}^\prime(t) + \bm{\omega} \times \left(\mathbf{R} \mathbf{x}^\prime(t)\right)
\end{align}
The rotation matrix $\mathbf{R}$ is:
\begin{equation}
  \mathbf{R} = \mathbf{R}_z({\bm{\theta}}_3) \mathbf{R}_y({\bm{\theta}}_2) \mathbf{R}_x({\bm{\theta}}_1) 
             %% = \left( \begin{array}{ccc} 
             %%     \text{cos}(\theta_2)\, \text{cos}(\theta_3) & 
             %%     -\text{cos}(\theta_2)\, \text{sin}(\theta_3) & 
             %%     \text{sin}(\theta_2) \\
             %%     \text{cos}(\theta_1)\, \text{sin}(\theta_3) + \text{sin}(\theta_1)\, \text{sin}(\theta_2)\, \text{cos}(\theta_3) &
             %%     \text{cos}\theta_1)\, \text{cos}(\theta_3) - \text{sin}(\theta_1)\, \text{sin}(\theta_2)\, \text{sin}(\theta_3) &
             %%     -\text{sin}(\theta_1)\, \text{cos}(\theta_2) \\
             %%     \text{sin}(\theta_1)\, \text{sin}(\theta_3) - \text{cos}(\theta_1)\, \text{sin}(\theta_2)\, \text{cos}(\theta_3) &
             %%     \text{sin}(\theta_1)\, \text{cos}(\theta_3) + \text{cos}(\theta_1)\, \text{sin}(\theta_2)\, \text{sin}(\theta_3) &
             %%     \text{cos}(\theta_1)\, \text{cos}(\theta_2) 
             %%   \end{array} \right)
\end{equation}
where $\mathbf{R}_x$, $\mathbf{R}_y$, and $\mathbf{R}_z$ are the standard rotation matrices about 
the $x$, $y$, and $z$ axes, and $\bm{\theta} = \bm{\omega} t$.

%==========================================================================
% Problem Description and Software Implementation
%==========================================================================

\section{Problem Description and Software Implementation}
\label{sec:implementation}

In this section we describe our white dwarf merger software, and focus in 
particular on the initial white dwarf models (\autoref{sec:initial_models}), 
the initial problem setup (\autoref{sec:initial_state}), and analysis 
(\autoref{sec:analysis}) components.

The software used to generate the test problems in this paper
(as well as the manuscript itself),
\wdmerger\footnote{\wdmerger\ can be obtained at \url{https://github.com/BoxLib-Codes/wdmerger}.},
is freely available at an online repository hosting service.
Version control in both the parent software (\boxlib, \castro) and in \wdmerger\
permits us to reference the state of the code at the time a simulation
was performed. In all plot files and diagnostic output generated by \castro, 
and figure files generated by \wdmerger,
we store the active \texttt{git} commit hashes of \boxlib, \castro, and \wdmerger.
Line plots are generated using the \matplotlib\ library for \python\ 
\citep{matplotlib}, while slice plots and other multi-dimensional visualizations are 
generated using the \yt\ code \citep{yt}.

\subsection{White Dwarf Models}
\label{sec:initial_models}

At the start of any full simulation, we generate initial model white
dwarfs by integrating the equation of hydrostatic equilibrium, taking
the temperature to be constant, and using the
stellar equation of state.  This results in a single non-linear
equation to find the density in a zone given the conditions in the
zone beneath it:
\begin{equation}
\frac{p_{i+1} - p_i}{\Delta x} = \frac{1}{2} (\rho_i + \rho_{i+1}) g_{i+1/2}.
\end{equation}
This equation is a function of $\rho_{i+1}$ only since the pressure is
uniquely determined by the density in this case. Here, $\rho_i$ and $p_i$
are known, and $g_{i+1/2}$ is the gravitational acceleration at the
interface between zones $i$ and $i+1$, found by simply adding up all
the mass from zones $1$ to $i$ to get the enclosed mass,
$M_{i+1/2}$, and then setting $g_{i+1/2} =
-GM_{i+1/2}/r_{i+1/2}^2$. We solve this equation for $\rho_{i+1}$
using a Newton-Raphson iteration.

We desire to specify the mass of the white dwarf, as well as its
temperature and composition. To start the integration off, we
therefore need to guess at a central density.  We then do a secant
iteration over the entire integration procedure to find the central
density needed to yield the desired total mass.  The grid spacing is
$\Delta x = 6.25\ \text{km}$. We chose this value because no simulation
we perform is likely to exceed this grid resolution inside the stars 
themselves; for our normal domain size (see below), this corresponds to 
three jumps in refinement by a factor of four. We find that for low 
resolution runs, this is a better choice than selecting the 1D grid 
spacing to be comparable to the 3D grid spacing.

The white dwarf composition is determined by the chosen mass. For 
this paper we adopt the scheme of \cite{dan:2012}. Low-mass WDs 
are pure helium; low-to-intermediate-mass WDs are an even carbon-oxygen 
core with a relatively large helium envelope; intermediate-mass 
WDs are a carbon-oxygen core with slightly more oxygen than carbon; 
and, high-mass WDs are composed of oxygen, neon, and magnesium. 
This choice of composition distribution broadly resembles the 
results of stellar evolution calculations in the respective 
mass ranges, though it does not match the calculations in detail.

We map the 1D model onto the 3D Cartesian grid by taking density,
temperature, and composition as the independent variables,
interpolating these to the cell centers, and then calling the equation
of state to initialize the remaining terms. It is possible to interpolate
instead by using pressure instead of temperature, as pressure is more 
closely related to hydrostatic balance, but the EOS we use is so 
insensitive to temperature that this mapping can result in large 
deviations from the isothermal assumption we started with.  The 
interpolation process divides each zone into $n_{\text{sub}}$ 
sub-zones of equal volume for
the purpose of sampling the 1D model, and the sub-zones are added
together to obtain the full zone's state. This
sub-grid-scale interpolation is useful especially near the edge of the star,
where the density falls off rapidly with radius. Typically we take 
$n_{\text{sub}} = 4$.

\subsection{Initial State}
\label{sec:initial_state}

For a single star simulation, the star is simply placed at the center
of the computational domain, which we take to be the origin. For a
binary star simulation, we take as parameters the mass of the two
white dwarfs and the initial orbital period $T$. Using Kepler's third
law and assuming a circular orbit, we can then work out the orbital
separation $a$:
\begin{equation}
  a = \left(\frac{GM T^2}{4\pi^2}\right)^{1/3}.
\end{equation}
Here $M = M_P + M_S$ is the total mass of the system, where $M_P$ is
the specified \textit{primary} mass and $M_S$ is the specified
\textit{secondary} mass. The primary WD always starts on the left
side of the computational domain for our simulations, and is more
massive than the secondary. This reflects the usual terminology in the
literature where the primary WD is the accretor and the secondary is
the donor. The center of mass is located at the center of the
computational domain, and by default the stars lie along the $x$ axis, so that
the primary's center of mass is located at $x = -(M_S / M)\, a$ and
the secondary's center of mass is located at $x = (M_P / M)\, a$.
The user may choose to initialize the stars along a different axis,
and can also choose a non-zero orbital phase and/or eccentricity.

The initial velocity is taken to be zero in if we are in the reference
frame that rotates with the WDs, and if we are in the inertial frame
the velocity in every zone is set equal to the rigid rotation rate 
corresponding to the distance of that zone from the rotation axis, given
the specified period $T$. Thus the inertial frame and rotating frame 
simulations are starting off with the same initial conditions: two white 
dwarfs locked in synchronous rotation. This is the simplest assumption to 
make, but in the future we may explore relaxing this requirement.

In the current paper we do not attempt to enforce equilibrium with an additional relaxation
step. This will be an important part of future work in this series, as
numerous groups working on binary evolution
\citep{swc:2000,motl:2002,rosswog:2004,dan:2011,pakmor:2012:gadget}
have commented on the importance of equilibrium initial conditions in
determining the evolution of the system. As a consequence of starting 
in a non-equilibrium setting, there are 
large density and pressure gradients near the white dwarf surfaces
that result in significant amounts of mass flowing out of the white dwarfs.
This can result in spurious non-physical consequences such as 
the total density or energy going negative in a zone. To compensate 
for this, we start the simulation with a timestep that is a few orders 
of magnitude smaller than that required by the CFL criterion, and allow
the timestep to increase by 1\% each timestep so that the timestep reaches 
its maximum allowed by the velocities on the grid over a span of approximately 
1000 timesteps. This allows the gas at the surface of the white dwarf
to come closer to equilibrium without having 
discontinuous jumps in the density or energy. For all simulations, 
the maximum timestep is set to be equal to one-half of the CFL limit.

The computational domain has a total size of $1.024 \times
10^{10}\ \text{cm}$ in each spatial dimension, and is centered at the
origin. Our coarse grid has $256^3$ zones, corresponding to a spatial
resolution of 400 km. For the present study, we choose a simple
refinement strategy: on the coarse grid, all zones within twice the Roche radius of each
star are tagged for refinement, using the formula provided by 
\citet{eggleton:1983} for the effective Roche radius $r_L$ of a star in a binary,
\begin{equation}
  \frac{r_L}{a} = \frac{0.49 q^{2/3}}{0.6^{2/3} + \text{ln}(1 + q^{1/3})}.
\end{equation}
In this formula we can use $q = M_S / M_P$ for obtaining the Roche radius of the secondary,
and use the inverse value of $q$ to obtain the Roche radius of the primary.
The extra buffer from doubling the Roche radius ensures that the sharp density gradients near
the edge of the star are within the zone of refinement. On higher levels, 
we tag all zones above a given density threshold (taken to be $1\ \text{g cm}^{-3}$ 
in this paper) that corresponds to the stars themselves. We also ensure
that the outer part of the domain is never tagged for refinement. In
future work we will add criteria that tag for refinement the gas
between the stars, which is expected to feature nuclear burning.

\begin{figure}
  \centering
  \includegraphics[scale=0.4]{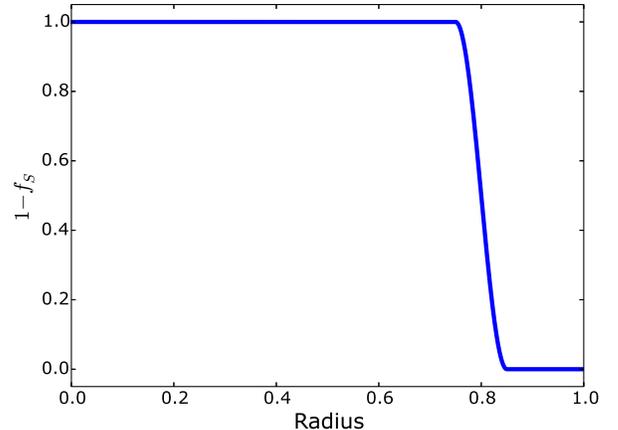}
  \caption{Radial profile of the hydrodynamical sponge we apply (\autoref{eq:sponge_frac}). 
           We subtract $f_S$ from unity; the value of $1 - f_S$ indicates what happens 
           to the sponged function after the sponge is applied. The sponge has no effect 
           in the inner part of the domain, and is fully applied at the outer edge.\label{fig:sponge}}
\end{figure}

Outside of the stars we fill the rest of the domain with a very low density 
ambient gas because our hydrodynamics model requires the density to be 
non-zero everywhere. This ambient material can create difficulties for the simulation.
In addition to the negative densities or energies at the stellar surfaces mentioned earlier, 
in the rotating reference frame we observe that standing instabilities can create very 
large velocities in the ambient fluid that drag down the global timestep by 
up to an order of magnitude.  To deal with this we employ a ``sponge'' similar 
to that described by \citet{maestro3} for the outer regions of the computational domain. 
After the hydrodynamics update, we apply a damping force to the momentum
equation as follows:
\begin{equation}
  (\rho \mathbf{u})^{n+1} \to \frac{(\rho \mathbf{u})^{n+1}}{1 + (\Delta t / \Delta t_S) f_S},
\end{equation} 
where $\Delta t_S$ is a timescale for the sponge to operate on, and
$f_S$ is the damping factor.  We choose it so that that the sponge is
non-operational inside a radius $r_S$ from the origin, and fully
applied at a radius $r_S^\prime \equiv r_S + \Delta r_S$. We then
smooth the sponge out between $r_S$ and $r_S^\prime$:
\begin{equation}
  f_{S} = \begin{dcases} 0 & r < r_S \\ \frac{1}{2}\left(1 - \text{cos}\left[\pi\left(\dfrac{r - r_S}{\Delta r_S}\right)\right]\right) & r_S \leq r < r_S^\prime \\ 1 & r \geq r_S^\prime. \end{dcases}\label{eq:sponge_frac}
\end{equation}
For the simulations in this paper we set $r_S$ to be 75\% of the 
distance from the origin to the domain boundaries, and $\Delta r_S$ so
that the sponge smoothing region extends another 10\% of that distance.
The resulting profile is displayed in \autoref{fig:sponge}. We set $\Delta
t_S = 0.01$ s, which is of the same order as the CFL timestep
for typical problem setups. While the sponge is applied we should avoid imputing any physical 
meaning to what is happening in the low-density gas far from the stars.

\subsection{Analysis}
\label{sec:analysis}

We track a number of diagnostic quantities at the end of coarse grid timesteps. 
For all simulations, we record the total energy (including the breakdown into
its components: kinetic, internal, gravitational potential, and rotation; we note
that for the diagnostics we actually use $(\rho E)$ for calculation of the total energy,
rather than explicitly calculating the sum of kinetic and internal, as this is
the quantity that should be explicitly conserved), 
the total angular momentum, and the center of mass of the system. 
We also separately record diagnostic 
information about the stars. Our strategy for tracking their 
locations is as follows: at the beginning of the calculation, we store the 
physical center of mass $\mathbf{x}_{c}$ of the stars as determined 
by Kepler's third law. We also store the velocity $\mathbf{v}_{c}$ 
of the stars. Then, at each new time step we make a preliminary guess for their 
location by updating the location using the old velocity, 
$\mathbf{x}_{c} \rightarrow \mathbf{x}_{c} + \mathbf{v}_{c} \Delta t$.
We then refine our guess for the location and velocity of each star by computing a
location-weighted sum of the mass and velocity over the computational domain. 
To do this, we need a cutoff for determining what counts as part of the primary 
and what counts as part of the secondary. We use a simple criterion: the star
that a zone ``belongs'' to is the one that exerts a larger magnitude
gravitational force on that zone (as computed using the tentative data
for that star's mass and radius). From this we obtain the corrected mass
of each star as well as its location and velocity. Once we have the new centers of mass,
we compute the effective radius of each star at various density cutoffs. This involves 
computing the volume $V$ of all zones that belong to the star (in the sense described above) 
whose density is greater than the cutoff. We then compute $r_{\text{eff}} = (3V/4\pi)^{1/3}.$

When we do simulations with adaptive-mesh refinement, there are multiple levels of refinement 
that contribute to a global integral. To deal with this we employ a ``mask'' which zeros out 
the data in a zone on a given level if there is a refined region overlying that zone.

\subsubsection{Gravitational Waves}
\label{sec:grvatational_waves}

A final diagnostic quantity we consider is the gravitational wave emission by the 
binary system. White dwarfs are not strongly affected by general relativistic effects;
the orbital motions are much slower than the speed of light, and the relativity parameter 
$GM / c^2 R$, which measures the ratio of the Schwarzschild radius of a mass $M$ to the actual 
radius $R$ of the object, is much less than unity for a white dwarf. Thus at any given time the 
relativistic effects are negligible compared to the Newtonian gravity and so we do not 
directly include relativistic effects in computing the dynamical evolution of the system.
A white dwarf binary system does emit gravitational waves during its evolution; this energy loss 
is what drives the initial inspiral over very long timescales. Eventually it will drive the system 
to become dynamically unstable due to the Newtonian tidal forces alone, though once that period begins, 
the gravitational energy loss is inconsequential in affecting the dynamical evolution of the system. 
The frequency of the gravitational waves emitted by the white dwarf binary 
is similar to the frequency of the orbital motion, which is in the range 
10-100 mHz for our problem. This is well outside the range of currently existing 
gravitational wave detectors but is very well suited for proposed space-based detectors such as eLISA \citep{eLISA}.

We follow the prescription of \citet{blanchet:1990} for computing a gravitational wave 
signal for our simulation. At distances far from the 
gravitational wave source, we can consider the leading term in the gravitational 
wave signal:
\begin{equation}
  h^{TT}_{ij}(t,\mathbf{x}) = \frac{2G}{c^4 r}P_{ijkl}(\mathbf{n}) \ddot{Q}_{kl}(t - r/c).
\end{equation}
$h$ is the perturbation to the spacetime metric and is commonly called the \textit{strain}; 
for laser interferometers, it measures the relative change in the distance between mirrors. 
The ``TT'' superscript indicates that we work in the commonly used tranverse-traceless gauge.
This strain is measured at time $t$ and position $\mathbf{x}$ relative to the binary system.
$r\equiv |\mathbf{x}|$ is the distance from the observer to the binary system. The unit vector 
$\mathbf{n} \equiv \mathbf{x} / r$ then measures the direction of the outgoing wave with 
respect to the observer, and $P_{ijkl}(\mathbf{n})$ is an operator that projects a tensor 
onto the direction orthogonal to $\mathbf{n}$:
\begin{align}
  P_{ijkl}(\mathbf{n}) &= \left(\delta_{ik} - n_i n_k\right)\left(\delta_{jl} - n_j n_l\right) \notag \\
                      &- \frac{1}{2}\left(\delta_{ij} - n_i n_j\right)\left( \delta_{kl} - n_k n_l\right).
\end{align}
$Q_{kl}$ is the quadrupole moment tensor:
\begin{equation}
  Q_{kl} = \int dV \rho \left(x_k x_l - \frac{1}{3}\delta_{kl} \mathbf{x}^2\right).
\end{equation}
The argument $(t - r/c)$ indicates that to get the strain at time $t$ we evaluate the second derivative of the 
quadrupole moment at the retarded time $t - r/c$. In practice the retarded time is simply the simulation time
and the observer would see the gravitational waves after a time delay of order $r/c$.

Therefore the primary component of the calculation is the evaluation of the second time derivative of $Q_{kl}$.
Explicitly constructing a discretized form of this derivative, using the current state and the state at 
previous times, is undesirable because of the inherent imprecision (its accuracy depends on the size of the timestep),
in addition to the logistical challenges that may be implied by saving and using previous simulation states. 
\citet{blanchet:1990} provide a prescription for this time derivative purely in terms of the state at a given time:
\begin{equation}
  \ddot{Q}_{kl} = \text{STF}\left\{2\int dV \rho (v_k v_l + x_k g_l)\right\}.
\end{equation}
The symmetric trace-free (STF) operator is defined as:
\begin{equation}
  \text{STF}\left\{A_{ij}\right\} = \frac{1}{2}A_{ij} + \frac{1}{2}A_{ji} - \frac{1}{3} \delta_{ij} \sum_{k}A_{kk}.
\end{equation}

The strategy is then as follows. At the end of the coarse timestep, we first calculate $\ddot{Q}_{kl}$
using an integral over the domain. This quantity is independent of the observer. If we 
are using a rotating reference frame, we first convert velocities and positions back to the inertial 
frame before evaluating the integral. Then, 
we pick an observing location $\mathbf{x}$ relative to the domain, evaluate the projection operator, 
and then perform the relevant tensor contraction to determine the strain tensor. We can 
repeat this process for any number of observing locations at minimal cost, since the quadruple tensor 
only needs to be calculated once. Gravitational waves only excite modes orthogonal to their 
direction of travel. These are the ``plus'' and ``cross'' modes, $h_+$ and $h_\times$, named after 
the types of spatial distortions they exhibit. We calculate the signal at a distance $r$ along 
the $x$, $y$ and $z$ axes. For the latter, as an example, $h_{+} = h_{11} = -h_{22} \propto (\ddot{Q}_{11} - \ddot{Q}_{22})/2$ and 
$h_{\times} = h_{12} = h_{21} \propto \ddot{Q}_{12}$. All other entries vanish. By default we take $r = 10$ kpc; 
as shown by \citet{loren-aguilar:2005}, this is a typical distance scale over which an 
experiment such as LISA could detect a coalescing binary white dwarf system. 
The strain at any other distance is easily calculated and goes as the inverse of the distance.

%==========================================================================
% Numerical Test Problems
%==========================================================================
\section{Numerical Test Problems}\label{sec:Tests}

White dwarf merger simulations face a number of numerical difficulties that are
not present in single-degenerate Type Ia and core-collapse supernova
simulations. In \autoref{sec:gravity}, we discussed how the lack
of spherical symmetry necessitates a careful look at the gravity
solver. There are also hydrodynamical issues: the merger process will
result in substantial motion of stellar material across the grid. This
bulk motion presents an opportunity for advection errors to build up,
and is only partially mitigated by evolving the white dwarfs in a
co-rotating frame. It is therefore important to be aware of the
behavior of the code in such circumstances. The behavior of \castro\ for
many standard hydrodynamics test problems was detailed in the original
code paper \citep{castro}, and in the interest of brevity we do not
repeat them all here. Instead, we focus on a subset of problems that
highlight the special difficulties introduced in merger
simulations. These problems couple the hydrodynamics, gravity and
equation of state modules. We observe that while in most non-trivial
three-dimensional problems this creates a complexity that makes it
impossible to determine exact analytical solutions, it is
straightforward to devise problems for which certain global properties
should obey simple, expected behaviors. Where possible, these should
be quantified and a convergence study performed, and that is
be the focus of the current section.

\subsection{Maintaining Hydrostatic Equilibrium}\label{sec:HSE}

In \autoref{sec:initial_models} we describe the process by which
we generate initial stellar models. While the 1D models are in
hydrostatic equilibrium to within a small error, interpolation onto
the 3D Cartesian grid will introduce perturbations into the solution
\citep{zingale:2002}. Although we ensure that the initial models are
generated with the same equation of state and are at least as well resolved as
our finest grid, there is still be a hydrodynamical error associated
with the fact that the rectangular grid cannot faithfully represent a
spherical star. Additionally, the gravitational potential obtained by
the multigrid solver will differ slightly from the one assumed by the
initial model, and the operator splitting between the gravity and
hydrodynamics should also result in small errors. As a result, we
expect that the star will oscillate slightly about an equilibrium
point, but that the amplitude of this oscillation should decrease with
increasing resolution.

\begin{figure}
  \centering
  \includegraphics[scale=0.45]{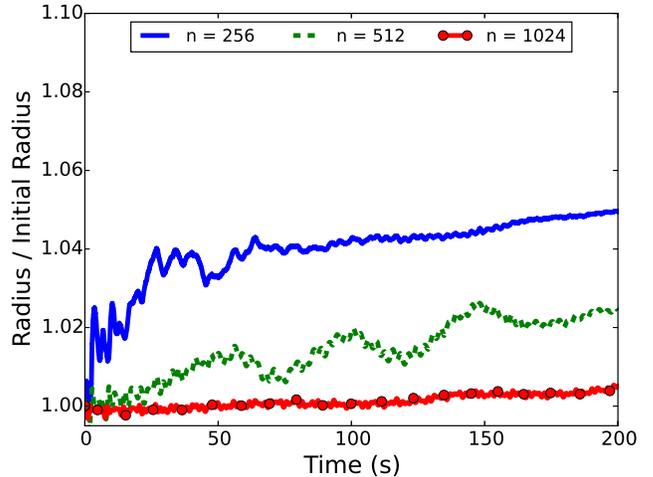}
  \caption{Time evolution of the effective radius of a $0.9 \msolar$ 
    white dwarf, seeded onto the grid using a one-dimensional hydrostatic
    model and evolved without further relaxation. The lines represent 
    different number of zones per spatial dimension; when this number is 
    greater than 256, it represents an effective resolution obtained 
    using AMR levels that cover the star. The radius is determined 
    using the volume of the grid that has a density greater than $10^3\ \text{g cm}^{-3}.$
    \label{fig:single_star_static_radius}}
\end{figure}

This problem was studied in the first \castro\ paper, but is worth
revisiting here. A single star explosion simulation may only last a
couple of seconds, and the \castro\ paper studied the behavior of the
star after one second of evolution. However, the dynamical timescale
of a typical carbon-oxygen white dwarf is on the order of 1--10
seconds. Additionally, a binary orbit is typically on the order of
10--100 seconds when a merger simulation starts, and with equilibrium
initial conditions the system may survive for tens of orbits before
the secondary is disrupted. When this does happen, we want to be
confident that it was because of the dynamics of the merger process
and not because of an instability in an individual star. Our goal here
is thus to install a single star onto our three-dimensional
coordinate grid and evolve it for a period of time long enough to
assess whether the star is truly stable, and to probe how the size of
deviation from equilibrium is affected by grid resolution.

\begin{figure}
  \centering
  \includegraphics[scale=0.45]{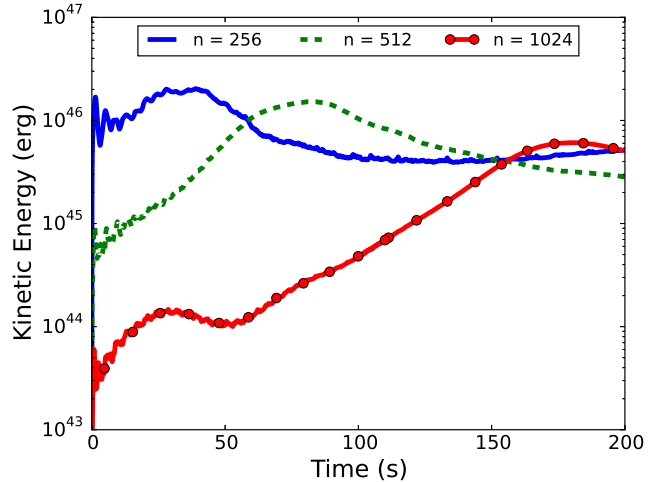}
  \caption{Time evolution of the kinetic energy of a $0.9\, \msolar$ 
    white dwarf. The lines have the same meaning as in \autoref{fig:single_star_static_radius}.
    \label{fig:single_star_static_ke}}
\end{figure}

We loaded a single star of mass $0.9\ \msolar$ onto the grid at the
origin, and evolved it for 200 seconds. Our diagnostic of choice is
the effective radius of the star, determined by the volume of the grid
that has a density greater than $10^3\ \text{g cm}^{-3}$ (see
\autoref{sec:implementation} for details on this measure). This choice
of density is intended to mark a reasonable outer edge to the star
that is not immediately susceptible to the numerical errors prevalent
near the physical edge of the star.
\autoref{fig:single_star_static_radius} shows our results at various
resolutions.  As expected, the star quickly approaches an equilibrium
size that is different (and in this case larger) than the
one-dimensional model, though the magnitude of this change becomes
smaller with resolution. The star is only approximately in equilibrium 
by this measure when the coarse grid of $256^3$ zones has a level of 
refinement that jumps by a factor of four. Even then there is a slight uptick
in the size toward the end, implying that the numerical stability is
not guaranteed for arbitrarily long timescales. For another view, we 
consider the kinetic energy on the grid, in \autoref{fig:single_star_static_ke}. 
This is a more holistic measure that weights the contribution by the density. 
At the end of the simulation the kinetic energy is not lower at the highest 
resolution than at the lower resolutions. This result suggests
that when constructing the equilibrium initial models that will form the
basis of later calculations, we should carefully monitor the evolution
of the stars when applying any artificial damping to cause the
merger, to ensure that the merger is due to this applied force and not
the intrinsic numerical instability of the stars.

\subsection{Gravitational Free Fall}\label{sec:Gravitational Free Fall}

\begin{figure}
  \centering
  \includegraphics[scale=0.45]{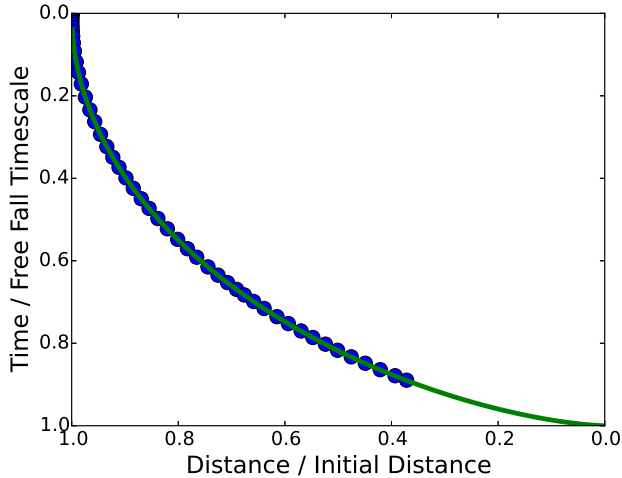}
  \caption{Time evolution of two initially stationary white dwarfs,
    mutually attracted to each other by the gravitational force. The
    horizontal axis gives the separation of the white dwarfs, scaled
    to the initial separation, and the vertical axis gives the elapsed
    time of the simulation, scaled to the time it would take two point masses
    to collide. The solid curve shows the analytical result,
    calculated from Newtonian mechanics, and the circles show the
    samples from the time evolution with \castro. For visual clarity, we 
    show only a small fraction of the timesteps.}
  \label{fig:freefall}
\end{figure}

A simple dynamical test to verify the coupling between the gravity and hydrodynamics in \castro\ is
the case of gravitational free fall. We place two stars on the grid 
in the manner of \autoref{sec:implementation}. The distance $a$ between 
them corresponds to a chosen orbital period $T$, consistent with the total
system mass $M$, but we disable the rotational source terms so that 
the stars start at rest in an inertial reference frame. 
Thus the stars will simply begin moving toward each other.
As long as the stars remain approximately spherical, the stars can be 
treated as point masses (this approximation only seriously breaks down after the stars
have come into contact). In dimensionless units where $r \to r / a$ and 
$t \to 2\sqrt{2}\pi t / T$, the simple free fall equation of motion governing the
distance $r$ between their centers of mass takes the form:
\begin{equation}
  \ddot{r}(t) = - \frac{1}{2r^2}.
\end{equation}
It is possible to derive a closed-form solution for the evolution time
as a function of separation by starting with the integral formulation,
\begin{equation}
  t(r) = \int_{1}^{r} \frac{dr}{v(r)}.
\end{equation}
The velocity $v$ (in dimensionless units) can be found by noting that 
$\ddot{r} = v\, dv / dr$ and then separating and integrating the equation 
of motion. This yields 
\begin{equation}
  v(r) = \sqrt{\left(\frac{1}{r} - 1\right)}.
\end{equation}
For our problem $0 < r \leq 1$, so this is always valid. Integrating, we find
\begin{equation}
  t(r) = \text{arccos}\left(\sqrt{r}\right) + \sqrt{r \left(1 - r\right)}. \label{analyticalFreeFall}
\end{equation}
so that the point of contact would occur at $t = 1$. We actually stop the simulation
at $t = 0.9$, which is when the effects from the extended sizes of the stars
starts to become important. The results of our simulation for our default $256^3$ zone 
uniform grid are shown in \autoref{fig:freefall}. They show excellent agreement
between the analytical solution and the simulation results.

\subsection{Galilean Invariance}\label{sec:galileo}

It is often stated in the literature that Eulerian methods for
hydrodynamics with grids fixed in space do not obey the Galilean
invariance of the underlying Euler equations, so that simulations
moving at a uniform bulk velocity appear different than an
equivalent stationary simulation (e.g. \cite{arepo}). If true, we need to understand 
the importance of this effect when deciding whether to trust the 
output of a code like \castro\ when applied for merger problems.
Recently, concern for the issue of Galilean invariance has come up in two ways which are of note for us 
in the present study. We explain these situations and display 
the results of tests we have run to determine whether this 
actually is a significant concern for our study.

\citet{arepo} (hereafter, S10) performed a Kelvin-Helmholtz instability test and showed
that (at low resolution) a fixed-grid code failed to develop the
expected fluid instability when the whole fluid was moving at a
strongly supersonic uniform velocity. (See also \citet{wadsley:2008}, 
who used the FLASH code to simulate a hot bubble subject to mixing 
by the Kelvin-Helmholtz instability, and also found that the mixing was affected by a 
uniform bulk velocity.) This contrasted with the results
of the moving-mesh code AREPO being presented in that study, which
demonstrated Galilean invariance even at large bulk velocities. 
Inability to correctly model the Kelvin-Helmholtz instability would 
have important consequences for how much we can trust the ability of 
\castro\ to test the violent merger progenitor model, where a detonation 
arises in the low-density material at the stellar surface. Shearing between 
the material flowing out of the secondary and material near the 
surface of the primary may trigger fluid
instabilities that play an important role in the evolution of that
gas, which is the site of the initial detonation in the prompt
explosion model. \citet{guillochon:2010} showed for their simulation
that Kelvin-Helmholtz instabilities produced this way may raise the
temperature of the accreting material enough to ignite a
detonation. Therefore if we are not correctly reproducing the
characteristics of the Kelvin-Helmholtz instability in the case where
there is significant mass motion on the grid, we cannot be confident
that a detonation (or lack thereof) is not numerically
seeded. 

\begin{figure*}[ht]
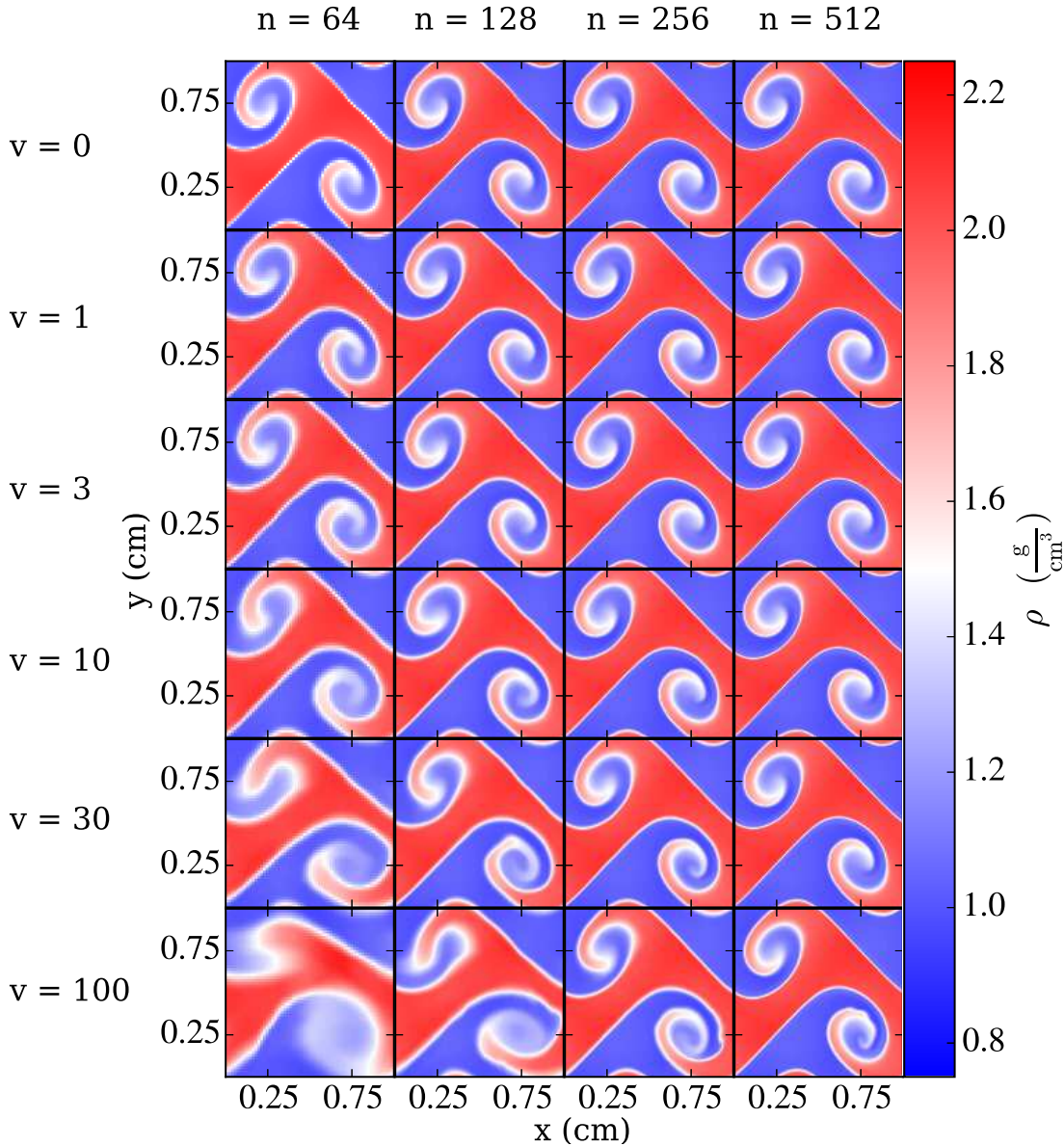

  % Note that since we have periods in the filename, we need to double-
  % bracket the filename so that includegraphics doesn't think it's 
  % looking for something with the wrong extension. See:
  % http://tex.stackexchange.com/a/10575
  \centering
  \includegraphics[trim=0.75in 0.45in 0.85in 0.5in, clip, scale = 0.75]{{{kh_t2.0_p2_low_res_collage}}}
  \caption{2D Kelvin-Helmholtz instability test at $t = 2.0$ for the initial 
    conditions given by \autoref{eq:kh_ic_b_ramp} and \autoref{eq:kh_ic_b}. 
    The rows each represent a different bulk fluid velocity $v$ and 
    the columns each represent a grid resolution $n$ (the number of 
    zones per spatial dimension). The highest velocity simulation, 
    $v=100$, corresponds to approximately Mach 70. Compare to 
    Robertson et al. (2010), Figure 7. \label{fig:kh_p2}}
\end{figure*}

\citet{robertson:2010} (hereafter, R10) observe that violation of Galilean
invariance of simulation results for the Euler equations occurs 
because of truncation error in the discretization of the fluid
equations. This takes the form of a numerical diffusion term which is
dependent on velocity (and also resolution). The advantage of a
moving-mesh code is that the mesh everywhere moves with the local flow
velocity, which substantially reduces the numerical
diffusion. R10 argue that the differences seen
between the moving-mesh and fixed-grid code are caused by the
interaction of this numerical diffusion with small-scale instabilities
(that may be physical or numerical) which couple with and
fundamentally alter the large-scale modes. Small-scale instabilities
are seeded by the choice of a sharp initial discontinuity between the 
fluids in the problem posed by S10. Crucially though,
R10 point out that this problem does not
converge with resolution (because the initial perturbation is too sharp 
and seeds numerical noise at the grid resolution level) 
and so it is not possible to know the correct
behavior of this problem. As such, we do not know whether the
small-scale modes found in AREPO are real, and the problem is not
useful in formally discriminating between methodologies. They instead
propose an alternate test with a smoother initial contact. This
converges to the same solution qualitatively in both the stationary
and bulk velocity cases, indicating that the code does generally
maintain Galilean invariance (to some specified error that depends on
resolution and the uniform flow speed).  We will see whether we can
reproduce this result.

A related question is whether our code reliably simulates the bulk
motion of the stars across the grid, and whether such bulk motion
affects the stability of the star. This concern is prompted by the
study of \cite{tasker:2008}, who studied the effect of uniform
translation on the stability of a spherically symmetric model for a
galaxy cluster. They compared the radial profile of the cluster at
initialization and after a period of time evolution. Using FLASH and
ENZO, they found that a static cluster retains its shape at high
enough resolution, while uniform translation of the cluster causes
mixing of the core material due to numerical diffusion which results
in an underestimation of the core's true density. The SPH codes they
used did a better job maintaining the core density. We will perform a
variant of this test using white dwarf models.

\subsubsection{Kelvin-Helmholtz Instability}\label{sec:khi}

Following \cite{robertson:2010}, we set up a Kelvin-Helmholtz test in
the following way. The problem domain runs from 0 to 1 in both the $x$
and $y$ directions. This is a two-dimensional test, so we run
\castro\ in 2D mainly to avoid extra computational expense; in 3D, it 
would merely involve replicating the problem in the $z$ direction.
The problem involves a fluid slab of density $\rho_2 = 2.0$ traveling rightward in the
$x$-direction at velocity $v_2 = 0.5$, sandwiched by a fluid of
density $\rho_1 = 1.0$ traveling leftward at velocity $v_1 =
-0.5$. The density gradient is in the $y$ direction, so this creates a
velocity shear along the interface between the fluids. The density and
velocity distribution on the computational domain are given by:

\begin{align}
  \rho &= \rho_1 + R(y)\left[\rho_2 - \rho_1\right] \\
  v_x  &= v_1 + R(y)\left[v_2 - v_1\right] \\
  v_y  &= v_{\text{bulk}} + v^\prime
\end{align}

Here $R(y)$ is a ramp function that describes the transition between
the two fluids, while $v_{\text{bulk}}$ is the bulk motion of the
fluid in the $y$ direction and $v^\prime$ is the velocity perturbation
that seeds the instability. The problem is established for two
sets of initial conditions (ICs), which we follow
R10 in calling ICs A and B. They differ in
their ramp function ($R_A$ and $R_B$ respectively), as well as the
initial perturbation ($v^\prime_A$ and $v^\prime_B$ respectively), and
the frequency of the perturbation ($n_A = 4$ and $n_B = 2$):
\begin{align}
  R_A &= \begin{cases} 0 & |y - 0.5| > 0.25 \\ 1 & |y - 0.5| < 0.25 \end{cases} \label{eq:kh_ic_a_ramp}\\
  R_B &= \Big\{\left[1 + e^{-2(y-0.25)/\Delta y}\right]\left[1 + e^{2(y-0.75)/\Delta y}\right]\Big\}^{-1} \label{eq:kh_ic_b_ramp}
\end{align}
\begin{align}
  v^\prime_A &= w_0\, \text{sin}\left(n_A\, \pi\, x\right) \left\{e^{-(y-0.25)^2 / 2\sigma^2} + e^{-(y-0.75)^2/2\sigma^2}\right\} \label{eq:kh_ic_a}\\
  v^\prime_B &= w_0\, \text{sin}\left(n_B\, \pi\, x\right). \label{eq:kh_ic_b}
\end{align}
Here $w_0 = 0.1$ is the scale of the velocity perturbation, $\sigma =
0.05/\sqrt{2}$ controls the width of the Gaussian for IC A, and
$\Delta y = 0.05$ is the transition distance scale for the smooth ramp of IC
B. The pressure everywhere is set to $p = 2.5$, and we run this with a
gamma-law equation of state set to $\gamma = 5/3$. Plotfiles are
generated every 0.05 seconds, and the problem is run until $t = 2$.

\begin{figure*}[ht]
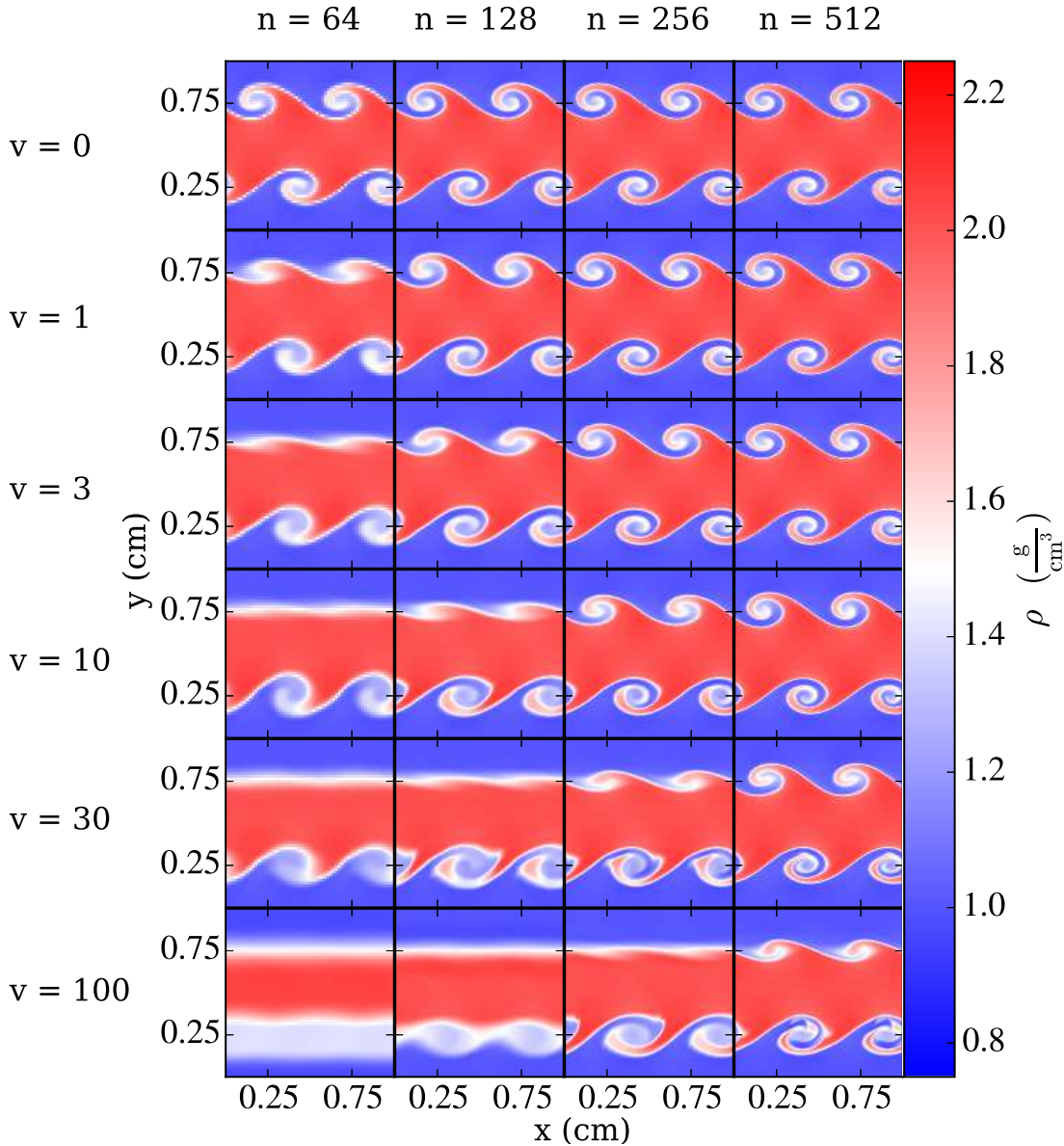

  \centering
  \includegraphics[trim=0.75in 0.45in 0.85in 0.5in, clip, scale = 0.75]{{{kh_t2.0_p3_low_res_collage}}}
  \caption{2D Kelvin-Helmholtz instability test at $t = 2.0$ for the initial 
    conditions given by \autoref{eq:kh_p3_rho} through \autoref{eq:kh_p3_vy},
    which come from McNally et al. (2012).
    The meaning of the rows and columns is the same as in \autoref{fig:kh_p2}.
    \label{fig:kh_p3}}
\end{figure*}

We run the problem for $v_\text{bulk} = [0, 1, 3, 10, 30, 100]$, and
for each set of initial conditions run the problem at resolutions of
$64^2$, $128^2$, $256^2$, $512^2$. For context, in these units the 
sound speed is $c\approx 0.7$. In addition, for each initial
condition we run simulations at the higher resolutions of $1024^2$,
$2048^2$, and $4096^2$ for the stationary problem only. These serve 
as a reference solution to gauge the extent to which the bulk flow 
affects the development of the fluid instability, and to determine 
if the problem is numerically converged.

We find the same result as R10 for IC A, which is equivalent to the 
test proposed by S10: at low resolutions and high bulk velocity, the 
Kelvin-Helmholtz instability completely fails to develop. Furthermore 
the problem does not converge even qualitatively at the highest 
resolutions we used. Our results are very similar to Figure 3 of 
R10 so we do not show them here. For IC B, our results can be seen 
for the normal resolutions and all velocities in \autoref{fig:kh_p2}.
At low resolutions and very large bulk velocities, the fluid 
does get significantly disrupted by numerical error. This 
effect quickly converges away with resolution and qualitatively 
at $512^2$ resolution the solution is nearly identical to the stationary 
$v=0$ problem. We agree with R10 that this problem does converge 
with resolution and is not subject to numerically-seeded secondary 
instabilities at the stopping time. This is evident even at low resolutions
by examining the first row of \autoref{fig:kh_p2}.

\begin{figure*}[ht]
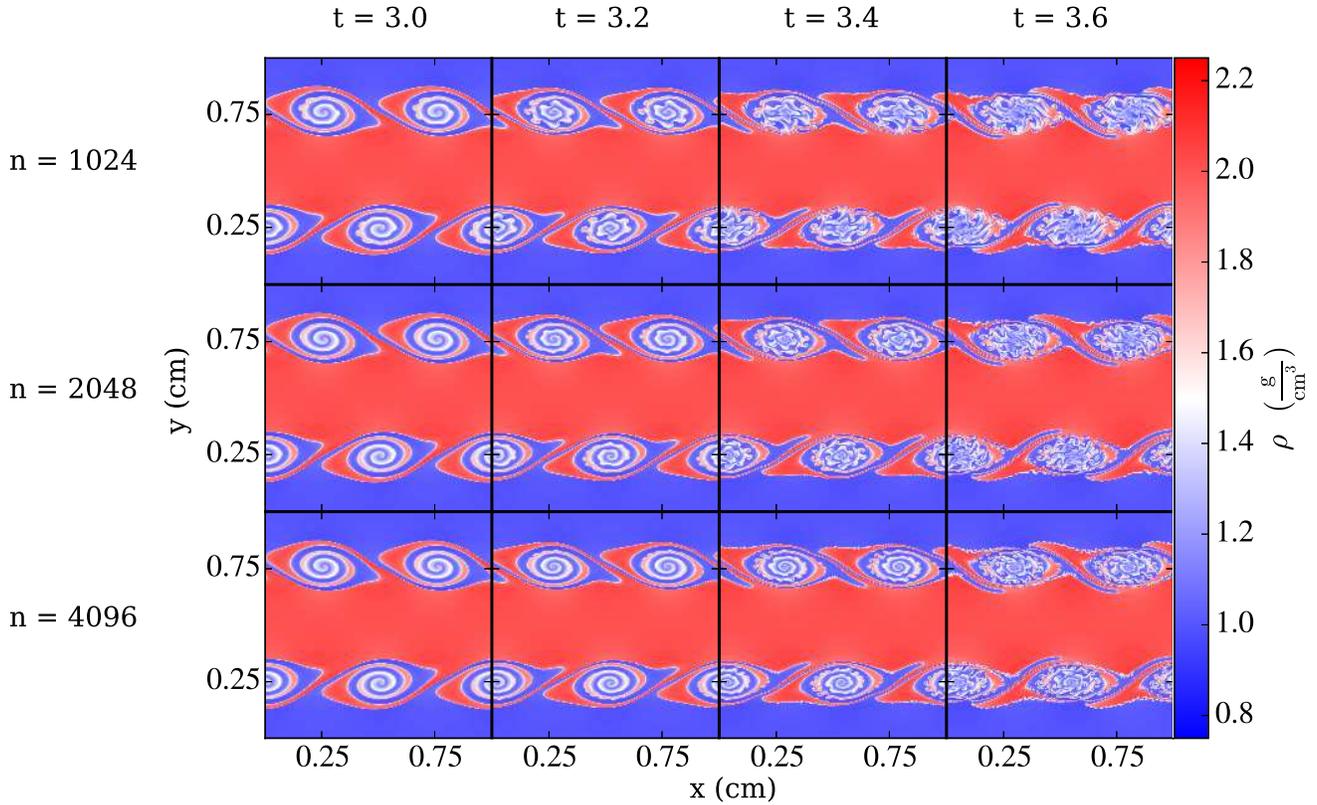

  \includegraphics[trim = 0in 1.25in 0.25in 1.25in, clip, scale=0.65]{{{kh_p3_high_res_collage}}}
  \caption{Time series of the Kelvin-Helmholtz problem proposed by McNally et al. (2012)
    as the simulation is just starting to go non-linear. The rows represent resolution, 
    where $n$ is the number of grid cells per spatial dimension, and the columns are 
    different snapshots in time.\label{fig:kh_p3_high_res}}
\end{figure*}

\citet{mcnally:2012} published another Kelvin-Helmholtz problem that 
is well-posed in the sense that it converges with resolution and 
is not subject to uncontrollable numerical instabilities. Though they 
were not explicitly interested in the question of Galilean invariance, 
we visit that issue here to see what can be learned. The initial 
conditions for this problem are:
\begin{align}
  \rho &= \begin{dcases} \rho_1 - \rho_m e^{(y-0.25)/\Delta y} & 0.25 > y \geq 0 \\ 
                         \rho_2 + \rho_m e^{(0.25-y)/\Delta y} & 0.5 > y \geq 0.25 \\
                         \rho_2 + \rho_m e^{(y-0.75)/\Delta y} & 0.75 > y \geq 0.5 \\
                         \rho_1 - \rho_m e^{(0.75-y)/\Delta y} & 1 > y \geq 0.75 \end{dcases} \label{eq:kh_p3_rho}
\end{align}
\begin{align}
  v_x &= \begin{dcases} v_1 - v_m e^{(y-0.25)/\Delta y} & 0.25 > y \geq 0 \\
                        v_2 + v_m e^{(0.25-y)/\Delta y} & 0.5 > y \geq 0.25 \\
                        v_2 + v_m e^{(y-0.75)/\Delta y} & 0.75 > y \geq 0.5 \\
                        v_1 - v_m e^{(0.75-y)/\Delta y} & 1 > y \geq 0.75 \end{dcases} \label{eq:kh_p3_vx} \\
  v_y &= w_0\, \text{sin}\left(4\pi x\right). \label{eq:kh_p3_vy}
\end{align}
Here $\Delta y = 0.025$, $w_0 = 0.01$, $v_m = (v_1 - v_2) / 2$, $\rho_m = (\rho_1 - \rho_2) / 2$, 
and the other symbols have the same meaning as above (this means the flow direction
is reversed compared to the original paper, so as to achieve consistency with the 
other simulations presented here). We run this problem at all the same resolutions 
and bulk velocities as the previous two problems. The results for the normal resolutions 
at $t = 2.0$ are displayed in \autoref{fig:kh_p3}. We see a similar pattern as for 
the test proposed by R10: as we get to higher flow speeds we need to have higher 
spatial resolution to compensate for the increased numerical diffusion. The 
qualitative accuracy is much lower for the highest bulk velocities for this problem
than for the previous problems. This is because the amplitude of the instability overall
is smaller than for the previous problems, at least by $t = 2.0$, so it is easier
for numerical diffusion at the shearing layer, caused by the high bulk velocities,
to completely wipe out the instability. Like \citet{robertson:2010} found for
their problem, we find for this problem that the convergence properties are not
substantially affected by altering the perturbation frequency -- the results
show the same qualitative pattern even if we halve this frequency.

\citet{hopkins:2015} performed this test as part of the testing of
their code GIZMO.  They showed the late-time evolution of this system,
when non-linear effects have taken over and significantly disrupted
the initial flow. At low resolution the tested grid algorithm
had failed to disrupt both for $v = 0$ and $v = 10$. We too ran this
problem until $t = 10$, and confirm that the Kelvin-Helmholtz
instability damps out at low resolution but goes strongly non-linear
and disrupts the flow at high resolution.  We strongly
emphasize the point that this does not objectively demonstrate a
deficiency in fixed-grid codes for this problem. We can only determine
the validity of a method when we have a trustworthy, converged
solution to compare to, and this is lacking for this problem at late
times. As observed by \citeauthor{mcnally:2012}, this lack of a solution is because the
secondary instabilities form for this problem when the whorls of the
Kelvin-Helmholtz tendrils stretch out and create gradients that
approach the grid resolution. This is prime breeding ground for
numerical noise. But because the nature of this noise depends on
the resolution, it is very different for simulations at different
resolutions. If these instabilities are seeded because of this
resolution-dependent noise and are not seeded instead in a controlled
manner such that they appear at the same time and location, then we
simply cannot draw any conclusions that bear on the question of
verification from this test at late times.
\autoref{fig:kh_p3_high_res} provides a sense of this by examining the
crucial time at which the transition from the linear to the non-linear
regime is occurring.  At all of these very high resolutions the
secondary instabilities develop, but they occur at different times and
have different spatial scales for each resolution.

We conclude that large bulk motions of fluid can have very significant effects 
on numerical calculations of shear mixing in fixed-grid codes, but that this effect 
diminishes with increasing resolution. As a result, we must be confident that we are 
sufficiently resolving the major mixing regions on the white dwarf surfaces,
specifically that the density gradients occur over spatial scales much larger
than the grid resolution. If we find instead that this mixing occurs near the
grid resolution scale, this will imply that we need to ramp up the resolution
in these regions using AMR. If this becomes too expensive, we would need to be
skeptical of any conclusions that could be drawn 
about the effect of the mixing on the nuclear burning.

\subsubsection{Moving Star}\label{sec:moving_star}

To analyze the effects of velocity-dependent results for a stellar simulation, we repeated the
test of \autoref{sec:HSE} with a bulk velocity on the grid. We chose a
velocity of $2.56 \times 10^{8}\ \text{cm s}^{-1}$. For context, this is 
comparable to the orbital velocities of the stars in \autoref{sec:kepler}, and the Mach
number is of order unity in the stellar core at this speed.  This test
was inspired by \citet{tasker:2008}, who considered a moving galaxy
cluster and who obtained a long timescale evolution by using
periodic boundary conditions, so that the cluster would cross the
domain multiple times throughout the evolution.  We believe that
periodic boundary conditions are unrealistic for our type of
simulation, so we prefer to do one continuous simulation where the
star does not cross the boundaries.  Since our normal grid was not
large enough to allow the motion to continue for very long, we
expanded the domain size by a factor of four, and then included an
extra refined level around the star to keep the effective resolution
the same. We started the star off in the lower left corner of the
domain, and pointed its velocity towards the upper right
corner. This allowed us to evolve the star for the same length of
time as for the original test. We note that getting the gravity
boundary conditions right required us to move the origin of the
problem at the bulk velocity, so that the multipole moments were
always computed with respect to the current location of the stellar center.

\begin{figure}
  \centering
  \includegraphics[scale=0.45]{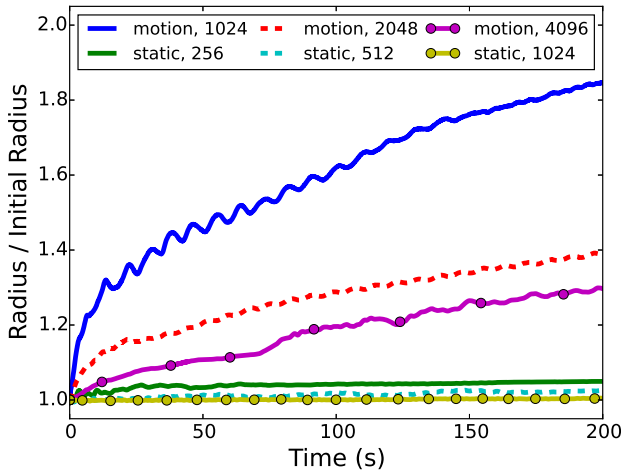}
  \caption{A variation on \autoref{fig:single_star_static_radius} where
    we now compare the ``static'' case to ``motion'' simulations where the 
    star moves across the grid at a fixed linear speed. The lines represent 
    the effective number of zones per dimension inside the stellar material;
    due to the expanded size of the grid in the ``motion'' case, the 
    physical resolution is the same in each column in the legend.
    \label{fig:single_star_compare_radius}}
\end{figure}

In \autoref{fig:single_star_compare_radius}, we take the results of
\autoref{sec:HSE} (the ``static'' case), and plot on top of it the
results of this new simulation (the ``motion'' case). We see
immediately that this bulk velocity causes the star to be much worse
at maintaining hydrostatic equilibrium. Not only is the absolute size
of the star significantly larger (nearly a factor of two at the lowest
feasible resolution we consider), but also there is a clear upward
trend in the size that has not terminated at any resolution by the end
of the simulation.  This again emphasizes the results mentioned
earlier, that we must be careful not to trust any simulation with
significant mass transfer if we are not confident that the mass
transfer is seeded in a controllable manner and free from numerical
noise.

\subsection{Keplerian Orbit}\label{sec:kepler}

We now consider the phase of the binary system where the stars are orbiting each other 
at distances great enough that the initial orbits should be approximately Keplerian. 
There are a number of effects worth looking into here. For simplicity, we choose two 
cases to demonstrate the simulation behavior: an equal mass case of two $0.9\ \msolar$ 
white dwarfs, and an unequal mass case of $0.9\ \msolar$ and $0.75\ \msolar$ white dwarfs.
In both systems, the secondary should be stable against mass loss.
In each case, the initial orbital period is 100 seconds.

\begin{figure}
  \centering
  \includegraphics[scale=0.45]{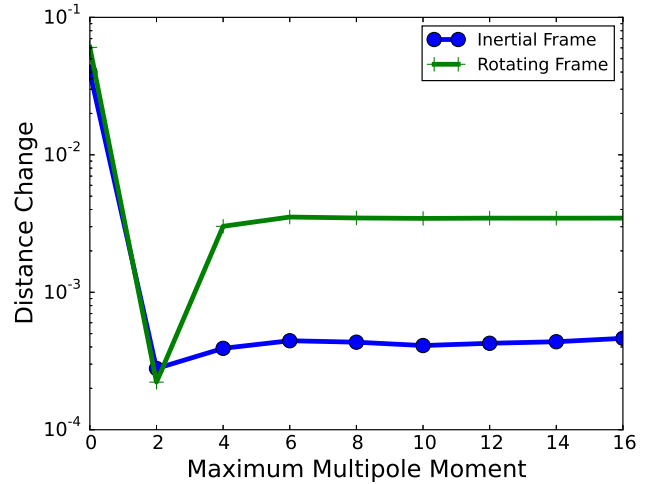}
  \caption{Absolute magnitude of the relative change in the distance of two unequal mass white dwarfs after one orbital period. 
           The stars were evolved in an inertial reference frame. The horizontal axis is the number of terms or multipole moments 
           captured in the series expansion for the potential at the domain boundary.\label{fig:gravity_bcs}}
\end{figure}

For some of the algorithms described earlier in this work, a single orbit of these 
systems is enough to examine their effects. In \autoref{sec:gravity_boundary_conditions},
we discussed the replacement of a monopole boundary condition solver for the gravitational 
potential with a more general multipole solver for the boundaries. To test the relevance 
of this effect, we considered a single orbit of the unequal mass system and measured 
the distance between the two white dwarf centers of mass at the beginning of the simulation and after 
the full orbital period. This distance should not change significantly over that timescale.
We performed this test for maximum multipole moments ranging from 0 (the monopole term) to 16.
The results are shown in \autoref{fig:gravity_bcs}. Terms in the boundary potential 
that vary faster than $r^{-5}$ are effectively negligible in determining the outcome of the orbit, 
justifying our typical choice of maintaining terms up to $r^{-7}$.

Another diagnostic that we consider is the energy conservation of the
system. Recalling \autoref{sec:gravity_hydro_coupling}, there are
several different methods of applying the gravitational source term to
the hydrodynamics equations. In \castro\ we presently have four
options, controlled by the parameter {\tt castro.grav\_source\_type},
which we shorten to {\tt gs} for the present discussion. 
${\tt gs} = 1$ and ${\tt gs} = 2$ are variations on the standard 
cell-centered source term for gravity. The difference between them is that 
${\tt gs} = 2$ determines the value of the energy source term after the momentum
source term has been applied, while ${\tt gs} = 1$ uses the uncorrected
momenta in calculating $\rho \mathbf{u} g$. We have found ${\tt gs} = 2$ to be
more accurate. ${\tt gs} = 3$ is entirely different: after calculating
the new momenta, we reset the total energy to be equal to the internal
energy plus the kinetic energy. This approach has the virtue of ensuring that
there is no conflict due to discretization between the momentum and
energy equations, and also correctly ensuring that the gravitational
force does not directly change the internal energy---and thus the
temperature---of the fluid. However, it explicitly sacrifices total
energy conservation. ${\tt gs} = 4$ is the new conservative method of
evaluating the energy source terms at cell faces. The results for the
change in energy after a single orbit are seen in the first column of
\autoref{table:sources}. The first two versions give reasonable and
similar levels of energy conservation. The third has total energy
changes on the order of 100\%, but this itself does not have a severe
effect on the dynamics because in this scheme the total energy
variable is effectively a placeholder value of the kinetic energy plus
internal energy, rather than being evolved directly. The last scheme
is nearly two orders of magnitude better in energy conservation,
justifying the effort in varying the scheme.

In \autoref{table:sources} we show also the effects on energy conservation of using the inertial reference frame. 
We use ${\tt rs}$ for the \castro\ parameter {\tt castro.rot\_source\_type}.
Each option for ${\tt rs}$ is implemented in the same way as for
the gravitational source term, simply swapping out the gravitational acceleration
for the rotational acceleration (except for the improvement to the momentum update
for ${\tt rs} = 4$ described in \autoref{sec:rotation}). 
The ${\tt rs} = 0$ column means that rotation is turned off and we are 
in the inertial frame. We see that the choice of rotational coupling is much less important than the choice of gravity coupling. 
The ``conservative'' ${\tt rs} = 4$ is slightly better in energy conservation than the non-conservative, 
cell-centered ${\tt rs} = 2$ algorithm, but it is a small effect.

\begin{deluxetable*}{llllll}
  \tablecaption{ Change in energy after a single orbit, i.e. $|\Delta E / E|$. ``rs'' is shorthand for the code parameter {\tt castro.rot\_source\_type} and ``gs'' is shorthand for the code parameter {\tt castro.grav\_source\_type}. The parameter meanings are explained in the main text.}
  \tablehead{ \colhead{        } &
              \colhead{ rs = 0 } & 
              \colhead{ rs = 1 } & 
              \colhead{ rs = 2 } & 
              \colhead{ rs = 3 } & 
              \colhead{ rs = 4 } } 
  \startdata
    gs = 1 & $ 4.8 \times 10^{-2} $ & $ 4.6 \times 10^{-2} $ & $ 4.6 \times 10^{-2} $ & $ 4.6 \times 10^{-2} $ & $ 5.7 \times 10^{-2} $ \\
    gs = 2 & $ 4.9 \times 10^{-2} $ & $ 4.6 \times 10^{-2} $ & $ 4.6 \times 10^{-2} $ & $ 4.6 \times 10^{-2} $ & $ 5.7 \times 10^{-2} $ \\
    gs = 3 & $ 1.1 \times 10^{ 0} $ & $ 2.8 \times 10^{ 0} $ & $ 2.8 \times 10^{ 0} $ & $ 2.8 \times 10^{ 0} $ & $ 2.8 \times 10^{ 0} $ \\
    gs = 4 & $ 4.4 \times 10^{-4} $ & $ 1.3 \times 10^{-3} $ & $ 1.3 \times 10^{-3} $ & $ 3.1 \times 10^{-4} $ & $ 1.0 \times 10^{-3} $
  \enddata
  \label{table:sources}
  % Castro git hash: 806b5c927e8116bd780bdff60164afef29acbf62
  % BoxLib git hash: e5d0548bbdfdca1aabd729f786b154f9c6340ce8
  % wdmerger git hash: 278a688171ee92ca04ce645a86518aabaf2391dd
\end{deluxetable*}

\begin{figure*}
  \centering
  \includegraphics[scale=0.7]{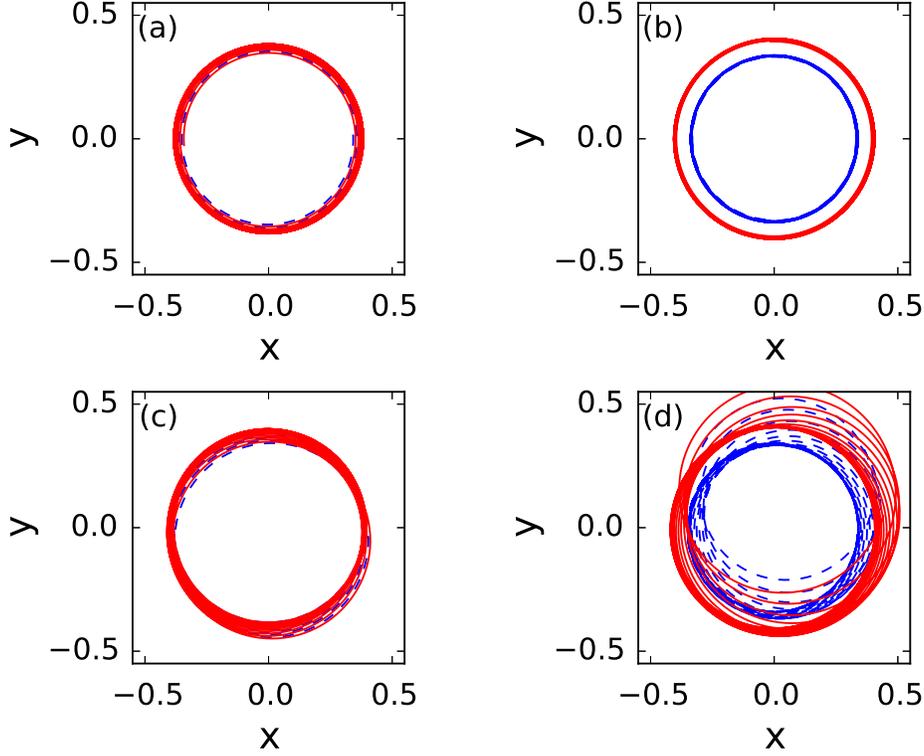}
  \caption{Positions of the white dwarfs in the orbital plane for four
    cases evolved over 25 orbital periods.  The $x$ and $y$ axes are
    normalized to the size of the domain, so that $x = -0.5$ is the
    left edge and $x = 0.5$ is the right edge. The dashed blue curve
    is the position of the primary white dwarf, and the solid red
    curve is the position of the secondary. In plot (a) we have the
    equal mass system evolved in the inertial reference frame, and in
    plot (c) we have the same system evolved in a rotating frame,
    where the positions have been transformed back to the inertial
    frame for comparison.  Plots (b) and (d) are analogous but for the
    unequal mass system.\label{fig:circular_orbit_comparison}}
\end{figure*}

We are most interested in the stability of these systems over long
timescales. To this end, we consider the same systems as above, but
evolve them for 25 orbital periods. In
\autoref{fig:circular_orbit_comparison} we illustrate the evolution of
these systems by plotting the center of mass locations of the white
dwarfs on the orbital ($xy$) plane. For the equal mass case in the
inertial reference frame, the curves fall nearly on top of each
other for most of the run, indicating that the stars are indeed 
orbiting at the initial distance, at least for a while. Towards the end 
of the run, however, the orbit starts to decay significantly, and the center-of-mass
distance of the two stars has decreased by about 10\% after 25 orbits.
We attribute this to non-conservation of angular momentum, which occurs 
because our code only explicitly conserves linear momentum. This orbital 
decay resembles the effect seen by \citet{swc:2000} for the case of neutron stars. 
In the unequal mass case, the magnitude of the orbital decay is smaller but 
at the end of the run the secular decline in distance is also visible. In 
both cases the stars would likely merge due to numerical error 
after a long enough timescale.

The co-rotating frame is different.  For clarity of visualization, we
rotate these results back into the inertial frame before displaying
their orbits.  In both the equal and unequal mass cases, the centrifugal force 
pushes the stars outward toward a new equilibrium distance that is a few 
percent larger than its initial distance. At the end of the run, the system is 
relatively stable, with oscillations about the new equilibrium distance. In fact 
these oscillations occur too in the inertial frame, but they are much more pronounced 
here. In the unequal mass case this is coupled with severe precession of the orbit, 
which results in chatoic-looking orbits when viewed from the rotating reference frame. 
These result from the explicit numerical consideration of the Coriolis and centrifugal 
terms, which do not appear in the inertial frame. So while the rotating frame 
is clearly more stable against mass transfer than the inertial frame,
the cost is that the specific dynamics may be more suspect.

\begin{figure}
  \centering
  \includegraphics[scale=0.45]{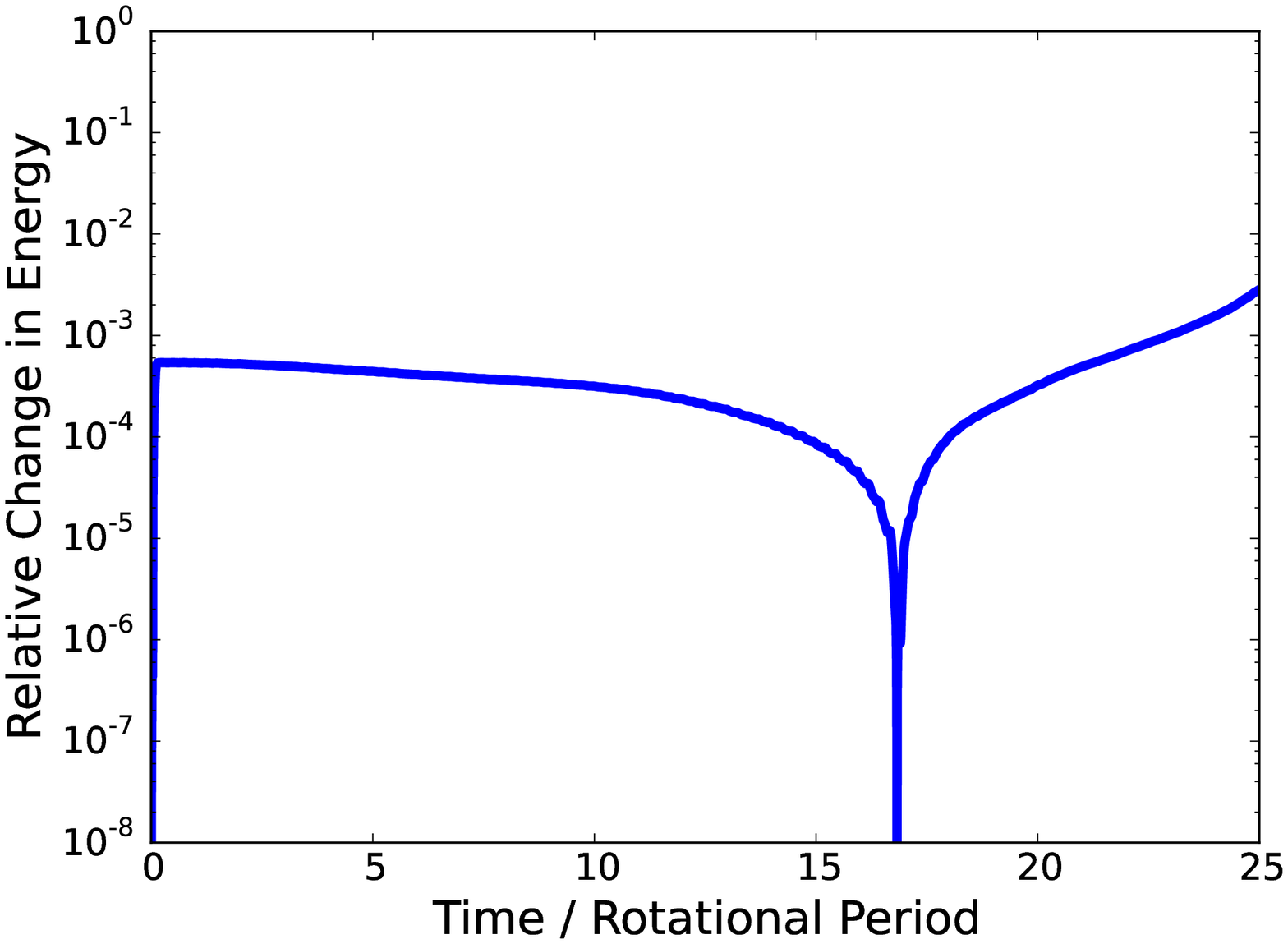}
  \caption{Absolute magnitude of the relative change in energy of two equal mass white dwarfs through 25 orbital periods,
           evolved in an inertial reference frame. The decline and recovery is a change in sign of the energy difference.
           \label{fig:energy_conservation_equal}}
\end{figure}

\begin{figure}
  \centering
  \includegraphics[scale=0.45]{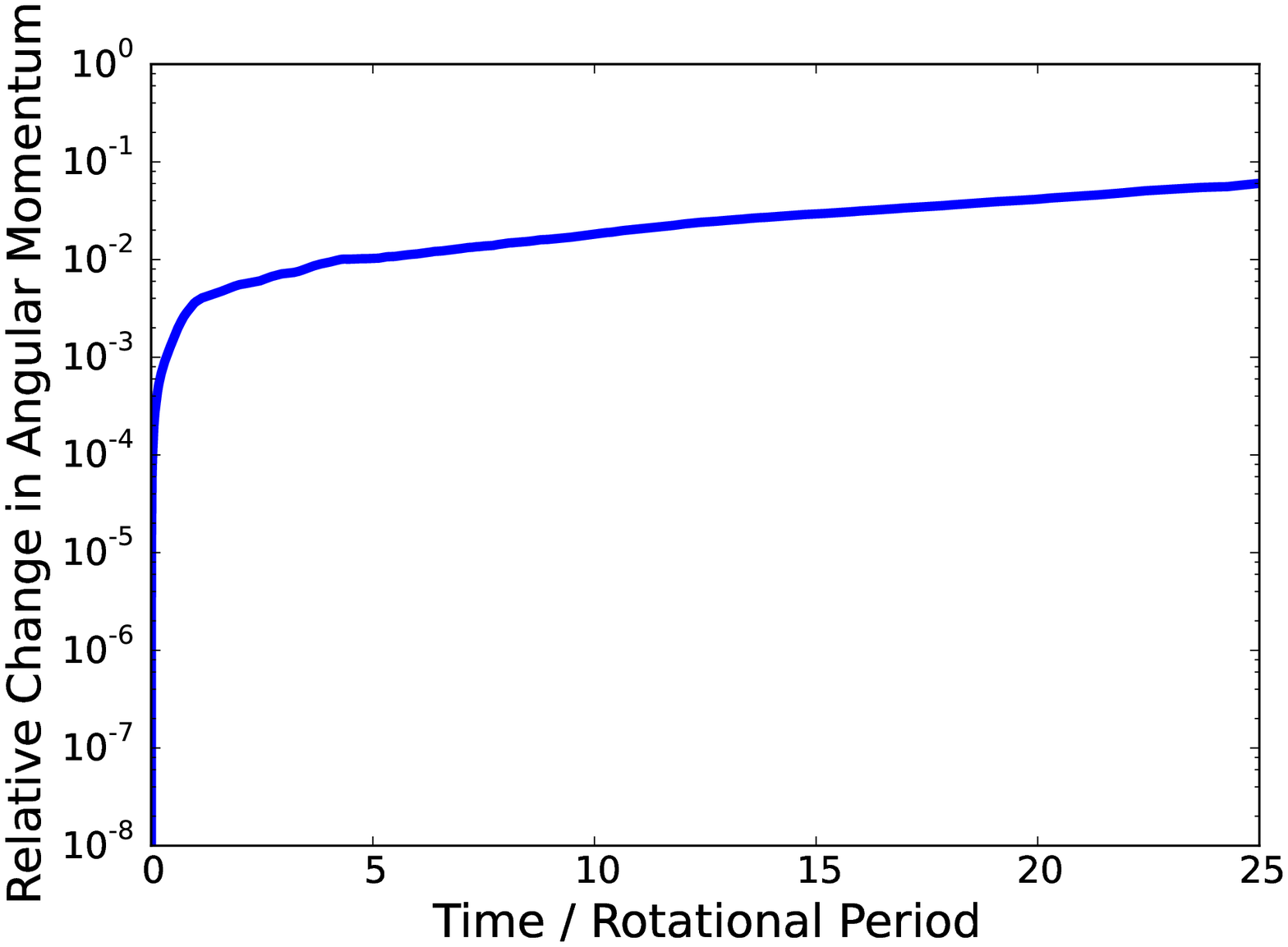}
  \caption{Absolute magnitude of the relative change in angular momentum of two unequal mass white dwarfs after 25 orbital periods, 
           evolved in a co-rotating reference frame. We consider only the component of the angular moment along the rotational 
           axis.
           \label{fig:angular_momentum_conservation_unequal}}
\end{figure}

Turning to the conservation properties of the system, we examine as
fairly typical cases the equal mass system in the inertial frame for
energy conservation (\autoref{fig:energy_conservation_equal}), and the
unequal mass system in the rotating frame for angular momentum
conservation (\autoref{fig:angular_momentum_conservation_unequal}).
For the former system angular momentum is conserved to within 10 percent over the 25
orbits, while energy conservation is about an order of magnitude
better. We note that while this is already a fairly good level of
energy conservation, it is not nearly as good as the results of
\citet{marcello:2012}. This is because we reset the internal energy to
a level corresponding to our temperature floor when it goes negative,
while \citeauthor{marcello:2012} do not reset and instead ignore the
internal energy if it is negative. The resets impose an artificial
floor on our ability to conserve energy, but they only happen in
low-density regions and do not much affect the large-scale dynamics.
Meanwhile, relative angular momentum conservation is not quite as good 
as relative energy conservation.  This is 
linked to the decline (or increase) in the size of the orbit. This
implies that we ought to be careful in concluding that at these
moderate resolutions we can safely evolve systems for many dozens of
orbits; this needs to be verified to ensure that an observed inspiral
and merger is physically (not numerically) motivated.

\begin{figure}
  \centering
  \includegraphics[scale=0.45]{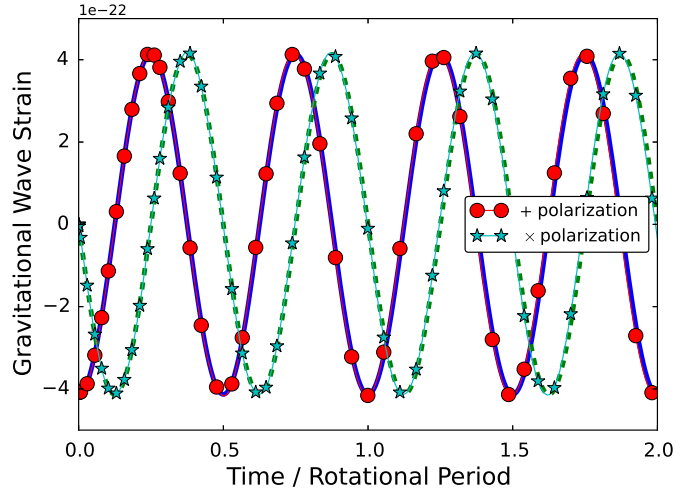}
  \caption{Gravitational wave strain polarizations for the first two orbital periods of an 
           unequal mass system. The curves with markers are the numerical data, while the 
           curves without markers are the analytical results for two point masses.\label{fig:gw_strain}}
\end{figure}

As a simple verification test to ensure our gravitational wave calculations are correct, we plot the 
gravitational wave strain along the rotation axis for the first two periods of an unequal mass system. 
At this early time the orbit is circular and so to a good approximation we expect that the gravitational 
wave signal should be that of two point masses, whose positions are:
\begin{align}
  \mathbf{r}_P(t) &= -a_P\, \text{cos}(\omega t) \hat{x} - a_P\, \text{sin}(\omega t) \hat{y} \\
  \mathbf{r}_S(t) &= a_S\, \text{cos}(\omega t) \hat{x} + a_S\, \text{sin}(\omega t) \hat{y}.
\end{align}
Then the mass distribution is $\rho(\mathbf{r}) = M_P\, \delta^3(\mathbf{r} - \mathbf{r}_P) + M_S\, \delta^3(\mathbf{r} - \mathbf{r}_S)$.
From this it is straightforward to calculate the quadruopole tensor, take its second time derivative, and then apply the 
projection operator to get the gravitational wave polarizations along the rotation axis:
\begin{align}
  h_+ &= -4\frac{G\mu}{c^4 r}\left[G M_{\text{tot}} \omega \right]^{2/3}\, \text{cos}(2\omega t) \\
  h_\times &= -4\frac{G\mu}{c^4 r}\left[G M_{\text{tot}} \omega \right]^{2/3}\, \text{sin}(2\omega t).
\end{align}
$\mu$ is the reduced mass, while $M_{\text{tot}}$ is the total mass. From this we see that the 
gravitational wave frequency is twice the orbital frequency, and that the two polarizations 
are out of phase by $90^\circ$ in time. We compare this analytical expectation to the 
numerical results in \autoref{fig:gw_strain}. We find very good agreement in this case, and this 
level of agreement holds in the rotating frame as well.

\begin{figure}
  \centering
  \includegraphics[scale=0.45]{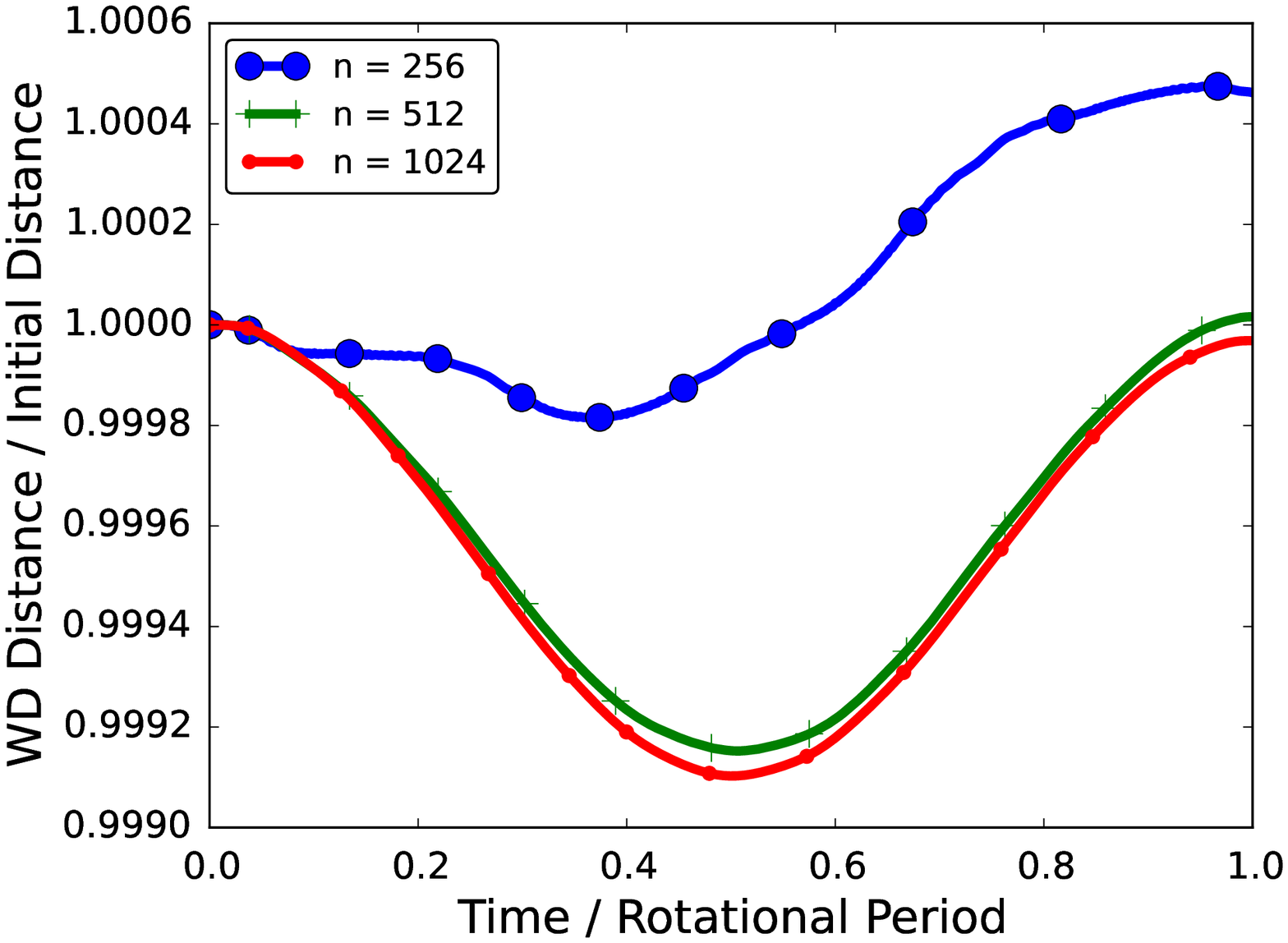}
  \caption{Distance between the two white dwarfs in the unequal mass system, for the first orbit.
           The distance is scaled by the initial orbital distance. 
           We plot at three different resolutions, corresponding to the number of 
           effective zones per dimension in the refined regions.
           \label{fig:unequal_spatial_convergence_inertial}}
\end{figure}

\begin{figure}
  \centering
  \includegraphics[scale=0.45]{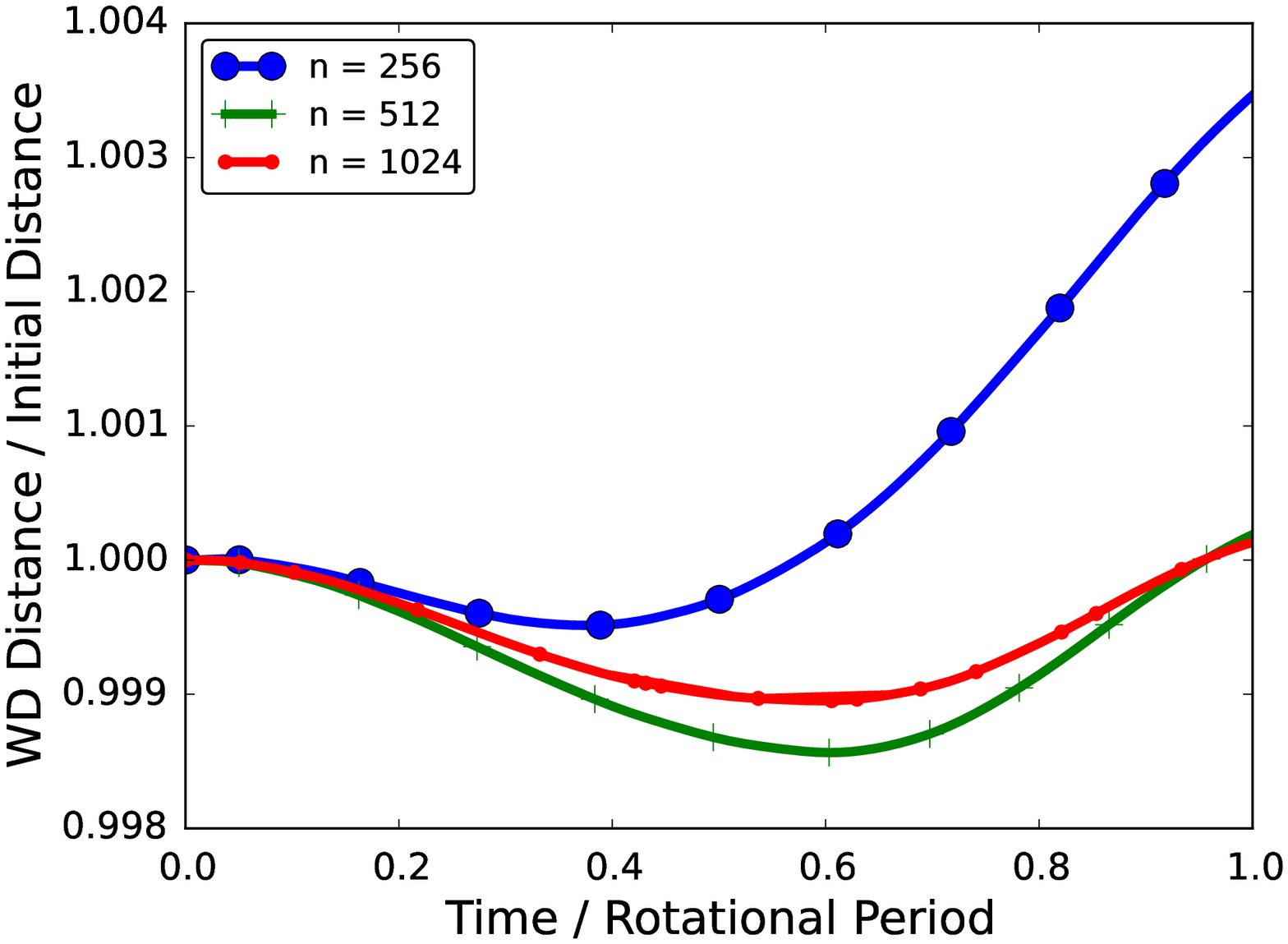}
  \caption{Distance between the two white dwarfs in the unequal mass system, for the first orbit.
           The distance is scaled by the initial orbital distance. 
           We plot at three different resolutions, corresponding to the number of 
           effective zones per dimension in the refined regions.
           \label{fig:unequal_spatial_convergence_rotating}}
\end{figure}

Finally we consider whether the dynamical behavior of the system converges with resolution. 
In \autoref{fig:unequal_spatial_convergence_inertial} we plot the first full orbit for 
the unequal mass system, at three different resolutions in the inertial frame: our default 
resolution of $256^3$ zones, as well as a single level of refinement with a jump by a factor of 
two (effective resolution $512^3$) or a jump by a factor of four (effective resolution $1024^3$). 
It is clear that at the latter resolution (corresponding to physical resolution of 100 km), 
we have achieved convergent behavior. In the rotating frame, the results also show
convergent behavior but the convergence is not as fast with resolution as in the inertial frame; 
see \autoref{fig:unequal_spatial_convergence_rotating}. At the two higher resolutions the white dwarf
distance is qualitatively similar, and both are qualitatively different from the lower resolution. However,
quantitatively the two higher resolution runs are not as similar to each other as the analogous runs in the
inertial frame. Convergence with resolution is slightly slower in the rotating reference frame
because in the rotating reference frame a stable, unchanging circular orbit requires balance between
two forces with opposite sign (the gravitational and centrifugal forces), and slight perturbations from the
circular orbit are amplified by the effect of the Corolis force. In the inertial frame, these numerical instabilities
vanish, but the cost is that there is no centrifugal force to actively maintain the white dwarf distance,
which is why it is much more likely for the orbit to prematurely decay. In either case, these results suggest
at least a minimum resolution of 200 km for getting the dynamics qualitatively right. To put that into context,
consider that the parameter study of \citet{dan:2014} used 40,000 SPH particles per simulation, or (for an equal mass
binary) 20,000 particles per white dwarf. For, say, a $0.9\ \msolar + 0.9\ \msolar$ white dwarf binary on
a $256^3$ zone simulation grid, there are 20,000 zones that fit within a white dwarf. We do not
intend here to directly compare results between the two simulation methods. We limit ourselves to the
observation that at least for grid-based codes, a parameter study such as the ones performed by
\citet{dan:2012} and \citet{dan:2014} would likely not yield qualitatively convergent results if it were to use the
same effective mass resolution. Instead the number of zones inside each star should at least be doubled.

%==========================================================================
% Performance
%==========================================================================
\section{Parallel Strategy and Performance}\label{sec:Performance}

\castro\ is designed to be deployed on high-performance computing systems using 
many thousands of processors simultaneously. It is worth briefly examining 
our strategy for parallelizing the problem over many computational nodes 
and our performance in situations similar to production science simulations. 
This is especially true because some aspects of our approach to parallelism 
have changed since the first \castro\ paper \citep{castro}, and we have obtained improved 
performance in certain settings.

The \boxlib\ framework that \castro\ is based on domain decomposes each AMR level into a number 
of boxes that collectively span the level. These boxes are distributed to processors 
through MPI parallelism; each MPI task in general holds multiple boxes and 
an update includes a loop over all the boxes an MPI task owns. The distribution 
obeys a load-balancing algorithm that attempts to equalize the amount of work 
done by each processor. \boxlib\ contains a number of strategies for distributing 
work in this way, and by default uses a space-filling curve approach with a 
Morton ordering (e.g. \cite{sasidharan:2015,beichl:1998}). By experiment we have found that the most efficient load-balancing 
strategy for our problem is actually a simple knapsack algorithm. In this approach, 
the amount of work owned by a processor is proportional to the number of grid cells 
associated with that processor, and the algorithm attempts to ensure that all 
processors have a similar number of total grid cells. We demand an efficiency of 0.9,
meaning that the average workload per processor should be no smaller than 90\% of the 
maximum workload found on any processor. We find that in practice the 
performance is largely insensitive to this choice.

The size and shape of grid boxes is an important consideration for efficiency. 
Boxes that are very small suffer from a host of problems, including the larger 
amount of communication required between hydrodynamics solves. Additionally, 
the multigrid solver is less efficient if the boxes are small because there 
are fewer available levels for coarsening and performing V-cycles. Conversely, 
boxes that are too large mean that there isn't enough work to go around when we 
have a large number of processors. Good performance is the result of a careful 
balance between these two effects. On the lower end, we require that all boxes 
be a multiple of 16 zones in each dimension; multigrid efficiency sharply decreases 
if this factor is any lower. On the upper end, we select the maximum grid size 
based on the number of processors we use and the total number of cells in the 
simulation. This size will therefore in general vary on different AMR levels. 
Generally we select a value in between 32 and 64 zones per dimension.

We use OpenMP to accelerate the work associated with the boxes owned
by each MPI task. Originally \castro\ used OpenMP to accelerate
individual loops in the hydrodynamics routines, such as the
piecewise-parabolic edge state reconstruction and the conservative
flux update. However, there is a significant amount of overhead
associated with generating a new OpenMP region at each of the many
different loops in a hydrodynamics algorithm. This makes such a
strategy sub-optimal for use on many-core processors and GPUs. We have
recently switched to a tiling approach where an OpenMP region is
generated at the start of the hydrodynamics routine and the individual
threads separately work on different partitions of each box \citep{boxlib-tiling}. This
results in much less overhead for the threading. In general we obtain
more efficient simulations than could be obtained using MPI only,
because there are fewer boxes and thus less communication for a given
number of processor cores. We are currently developing an approach to
evaluating the hydrodynamics and microphysics modules on GPUs,
which will allow us to
take advantage of the significant computational resources embedded in
GPUs on certain systems.

\begin{figure}
  \centering
  \includegraphics[scale=0.4]{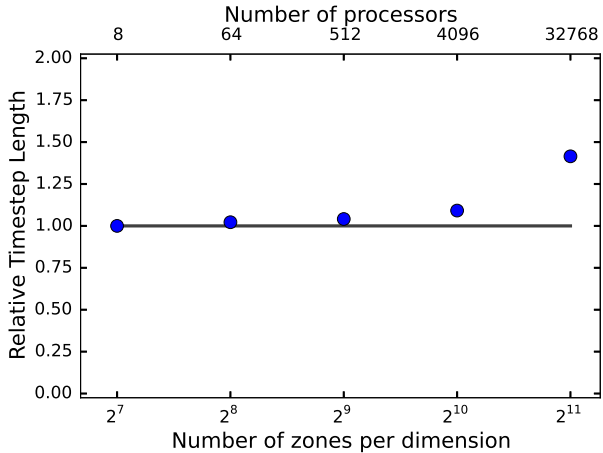}
  \caption{\castro\ weak scaling test, performed on Blue Waters at 
    NCSA. Each processor had a fixed amount of work, and we increased the 
    number of simulation zones in concert with the number of processors. The 
    solid curve represents perfect weak scaling, while the blue circles show 
    \castro's performance at each processor count. The vertical axis measures 
    the median time per timestep, normalized to this value for the smallest 
    processor count.\label{fig:weak_scaling}}
\end{figure}

To examine the parallel performance of \castro, we performed both
strong scaling and weak scaling tests on the Blue Waters machine at
the National Center for Supercomputing Applications. For the weak
scaling test, whose results are shown in \autoref{fig:weak_scaling},
we ran a uniform grid binary white dwarf simulation for resolutions of
$128^3$ zones through $2048^3$ zones. The number of processors was
scaled with the number of zones so that each processor had the same
amount of work; the smallest test used 8 processors and the largest
used 32,768 (note that the number of processor cores on a Blue Waters
node is twice the number of floating point units on that node). The
test was run for 10 timesteps, with each timestep including two
Poisson solves and a hydrodynamics update (though for a uniform grid
calculation we generally do not need to perform any multigrid
iterations for the first Poisson solve in a timestep, since the
density distribution has not changed since the end of the last
timestep). We disabled plotfile and checkpoint writing, as well as
calculation of diagnostic information (the latter can contribute to a
significant fraction of the run time at large processor counts if
computed every timestep). We computed the median wall time required
per time step for each simulation, and then normalized this to the
median time per timestep for the smallest simulation. We find
excellent weak scaling through 4,096 processors. At the largest run,
the simulation time required is slightly less than 1.5 times the amount
required for the smallest simulation.  This is due entirely to the
increased cost of the multigrid Poisson solve in each timestep and
this cannot be mitigated except by improving communication or
computation efficiency in the multigrid solver. We observe that this
weak scaling behavior with Poisson gravity is a significant
improvement over the results presented in the first \castro\ paper.

\begin{figure}
  \centering
  \includegraphics[scale=0.4]{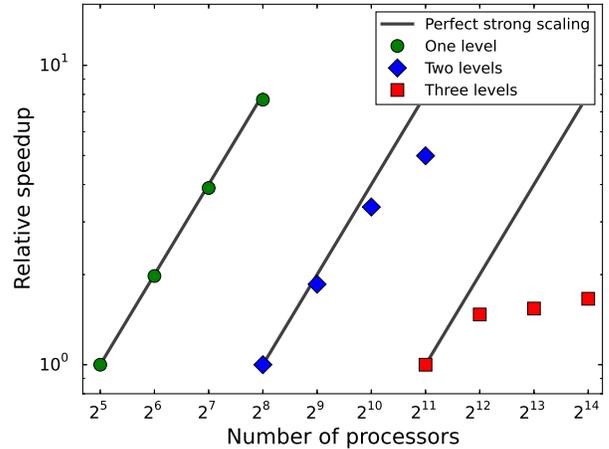}
  \caption{\castro\ strong scaling test performed on the Blue Waters machine at
    NCSA. The vertical axis measures the median time per timestep, and the 
    horizontal axis measures the number of processors in the simulation. Data 
    points are normalized to the time per timestep for the smallest number 
    of processors. The green circles show the data for a simulation with one 
    AMR level (a single uniform grid), the blue diamonds show the data for a simulation with two AMR
    levels (one coarse and one fine), while the red circles show the data 
    for a test with three AMR levels (one coarse and two fine). The fine levels 
    increase the resolution only in the regions around the stars. For each case 
    we draw a solid curve representing perfect strong scaling.\label{fig:strong_scaling}}
\end{figure}

The strong scaling test we perform uses a grid setup similar to what we 
use for well-resolved binary simulations. With only a uniform coarse grid, 
there are approximately $2 \times 10^7$ zones. With a single refined level, 
we have approximately $2 \times 10^8$ zones, typically spread over $\sim 2000$ grids.
On a second refined level, there are a similar number of zones and grids 
(the volume covered by this level is smaller, which offsets the greater resolution).
We run a scaling test for all three cases, with the 
highest processor count in each case chosen so that the number of 
MPI tasks is similar to the number of grids. There are no gains 
to be achieved from further parallelism. The results are found in 
\autoref{fig:strong_scaling}. We find excellent scaling for low to moderate numbers 
of processors. Parallel efficiency is well maintained when there are 
at least 2 grids per processor. The 
scaling behavior worsens at the highest processor counts, but this is 
an expected consequence of processors becoming work-starved. At the highest 
processor count in this test, there is approximately only one grid per processor. In general 
we find very good strong scaling behavior in the regime we are presently 
interested in, simulating the early phases of a simulation at moderate 
resolution. The strong scaling behavior is acceptable, though not perfect, 
at very large processor counts when self-gravity is considered.

%==========================================================================
% Conclusions
%==========================================================================
\section{Conclusions and Discussion}\label{sec:Conclusions and Discussion}

In this paper we have described the major components of a framework for 
simulating mergers of white dwarfs. While there is much evidence for the 
hypothesis that mergers (or collisions) of white dwarfs are significant 
contributors to the rate of Type Ia supernovae and related astronomical 
transients, the theoretical view of these systems is far from complete. 
Studying these systems over the long timescales relevant to dynamical 
mass transfer requires careful attention to the numerical methods used, 
to ensure that numerical instabilities or other errors do not unduly 
influence the system. Here we have described a number of common problems
that may occur, including violation of the conservation of energy, 
a lack of hydrostatic equilibrium (at low resolution) of the stars 
even when not acted on by external gravitational forces, and large 
velocities that can be generated near the edges of stars due to the 
numerically sharp gradients. Some of the issues are simply unresolvable
at the resolutions achievable on modern supercomputing systems; 
for example, it is very difficult to adequately resolve the stellar 
surface of a white dwarf on a three-dimensional grid, and there is 
justifiable room for suspicion regarding what happens there. But others
are avoidable with care: energy non-conservation can be substantially 
mollified by using a form of the gravitational work that is explicitly 
conservative, and we have observed that this can be done for rotation 
source terms too.

We presented a set of numerical tests that show where we can 
and cannot trust these techniques. We spent much time considering 
the role of bulk motions on the grid and we conclude that there 
are real issues with substantial bulk velocities on static grids, 
that can diminish the quality of the resulting solutions to the 
fluid equations, but that these effects diminish with increasing 
resolution. We therefore make no explicit claims about the usefulness 
of Langrangian versus Eulerian methods and instead simply observe 
that whenever a simulation is performed, it is important to have 
a measure of numerical accuracy and a sense of whether we are 
witnessing properties that converge with increasing resolution.
These problems do suggest that, where possible, we should seek to 
minimize bulk motions on static grids. A comparison of orbit 
simulations in both rotating and inertial reference frames 
demonstrates that in practice this is not so simple, and that 
a rotating reference frame has its own numerical issues for 
the type of simulation we desire to perform here. While decay 
of the stellar orbit is a commonplace feature in the inertial frame 
but easier to avoid in the rotating frame, the rotation forces 
can result in likely unphysical oscillations of the stars. It 
is not easy to predict the correct behavior of such systems,
and therefore determine which frame is closer to being correct, 
though one recourse for assessing confidence in a model from a 
verification standpoint is to see whether the observed 
behavior converges with resolution. It is not clear 
whether at practical resolutions the results in the 
rotating and inertial frames will converge to each other.

Future work on this project will focus on how to build 
reliable equilibrium initial models of the stars and 
an examination of how the mass transfer episode depends on 
these initial conditions, and then to enable the nuclear 
reaction network and determine whether self-consistent 
thermonuclear detonations are ignited and result in 
events that appear similar to Type Ia supernovae. Other 
areas ripe for future study include: the effect of radiation 
on the merger process; the extent to which the result 
depends on the initial composition of the stars (for example, 
by studying the dependence of the results on the size of helium 
surface layers on carbon-oxygen white dwarfs, or by using 
white dwarf models generated by modern stellar evolution 
codes); and, collisions of white dwarfs. All of 
these are possible under the framework we have established.

\acknowledgments

The authors thank John Bell, Volker Springel, and Dominic Marcello
for helpful discussions on gravity 
and hydrodynamics issues. We also thank Noel Scudder and Platon Karpov 
for their help with this project, especially related to visualization of the results, 
and Adam Jacobs for his advice on and assistance with 
running on supercomputing resources. Special thanks are given 
to the organizers of the 2015 Caltech Gravitational Wave 
Astrophysics School, and their supporters: the NSF 
under CAREER award PHY-1151197, the Sherman Fairchild 
Foundation, the LIGO Laboratory at Caltech, and Caltech.

This research was supported by NSF award AST-1211563. An
award of computer time was provided by the Innovative and Novel
Computational Impact on Theory and Experiment (INCITE) program.  This
research used resources of the Oak Ridge Leadership Computing Facility
located in the Oak Ridge National Laboratory, which is supported by
the Office of Science of the Department of Energy under Contract
DE-AC05-00OR22725. Projects AST006 and AST106 supported use of the ORNL/Titan resource. 
This research is part of the Blue Waters sustained-petascale computing project, 
which is supported by the National Science Foundation (awards OCI-0725070 
and ACI-1238993) and the state of Illinois. Blue Waters is a joint 
effort of the University of Illinois at Urbana-Champaign and its 
National Center for Supercomputing Applications.
This research used resources of the National Energy Research Scientific Computing
Center, which is supported by the Office of Science of the
U.S. Department of Energy under Contract No. DE-AC02-05CH11231.
Results in this paper were obtained using the high-performance
LIred computing system at the Institute for Advanced Computational
Science at Stony Brook University, which was obtained through
the Empire State Development grant NYS \#28451. 
This work used the Extreme Science and Engineering Discovery Environment (XSEDE), 
which is supported by National Science Foundation grant number ACI-1053575. 
Project AST100037 supported use of the resources NICS/Kraken and NICS/Darter.

This research has made use of NASA's Astrophysics Data System 
Bibliographic Services. In addition, this research has made use
of the AstroBetter blog and wiki.

\software{BoxLib (\url{https://github.com/BoxLib-Codes/BoxLib}),
          CASTRO (\citet{castro}, \url{https://github.com/BoxLib-Codes/Castro}),
          wdmerger (\url{https://github.com/BoxLib-Codes/wdmerger}),
          GCC (\url{https://gcc.gnu.org/}),
          python (\url{https://www.python.org/}),
          matplotlib (\citet{matplotlib}, \url{http://matplotlib.org/}),
          yt (\citet{yt}, \url{http://yt-project.org/})}

\clearpage

\appendix

\section{CASTRO hydrodynamics changes}
\label{app:hydro}

The basic PPM algorithm in \castro\ has undergone a number of changes
since the original code paper \citep{castro}.  A discussion of the
pure hydrodynamics changes along with verification of \castro\ when
using the stellar equation of state was given in
\citet{zingalekatz:2015}.  Here we discuss the changes that affect
multispecies flow and source terms.

\subsection{Reference States}

For all the runs, the PPM reconstruction is done using the original limiters for
the parabolic profiles \citep{ppm}; see \autoref{sec:Hydrodynamics} for a brief 
discussion about the limiters.  The prediction of the interface
states appears as:
\begin{equation}
\label{eq:ppmstatel}
q_{i+1/2,L}^{n+1/2} = \tilde{q}_L -
   \sum_{\nu;\lambda_i^{(\nu)}\ge 0} l_i^{(\nu)} \cdot \left [
        \tilde{q}_L  - \mathcal{I}^{(\nu)}_+(q_i)
       \right ] r_i^{(\nu)}
\end{equation}
where $q$ is the vector of primitive variables, $l_i^{(\nu)}$ and
$r_i^{(\nu)}$ are the left and right eigenvectors with eigenvalue
$\lambda_i^{(\nu)}$, with $\nu$ the index of the characteristic wave of
the system.  The sum is over all the waves that result from the
characteristic structure of the problem, but designed such that only
waves moving toward the interface contribute to the interface value,
$q_{i+1/2,L}^{n+1/2}$.  The reference state, $\tilde{q}_L$ is
chosen to minimize the work of the characteristic projection.
Finally, $\mathcal{I}_+^{(\nu)}(q)$ is the
average under the parabolic profile of quantity $q$ of all the
information that can reach the right interface of the zone $i$ as
carried by the wave $\nu$.  The reader is referred to
\citet{ppmunsplit} for further details.

Since the original \castro\ paper, the reference state implementation
has been switched to:
\begin{equation}
\label{eq:refchoice}
\tilde{q}_L = \left \{ \begin{array}{cc}
       \mathcal{I}_+^{(+)}(q_i) & \mathrm{if~} u + c > 0 \\
       q_i                    & \mathrm{otherwise}
\end{array}
\right .
\end{equation}
where the $(+)$ superscript here means the fastest wave moving to the right
(the $u+c$ eigenvalue).   This is simply the average under the largest
portion of the parabolic profile that could possible reach the interface 
over the timestep.  This is
in agreement with \citet{ppmunsplit} (eq. 90).  The flattening
in the original \castro\ paper has also been updated as discussed
in \citet{zingalekatz:2015}.

We comment on the choice of reference state for passively-advected
quantities (like $X_k$ or the transverse velocity), which is not typically 
discussed. First, consider one of the variables present in one-dimensional flow
(density, velocity in the normal direction, and pressure), and let our 
reference state be as in \autoref{eq:refchoice}.
Ignoring flattening, if there are no waves moving toward our
interface, then \autoref{eq:ppmstatel} reduces to:
\begin{equation}
q_{i+1/2,L}^{n+1/2} = \tilde{q}_L = q_i
\end{equation}
If instead only the fastest wave is moving toward the interface, then
only the term corresponding to the fastest wave in the sum will be
added in \autoref{eq:ppmstatel}, but our choice of reference state makes
that term zero by design, and our interface state is:
\begin{equation}
q_{i+1/2,L}^{n+1/2} = \tilde{q}_L = \mathcal{I}_+^{+}(q_i)
\end{equation}
This is the desired behavior for each of these cases. 

However, now consider the same approach applied to passively advected quantities.
If we use the same idea of the reference state as
in \autoref{eq:refchoice}, and consider a quantity $\xi$ which should only
jump across the contact, then our interface state becomes:
\begin{equation}
\xi_{i+1/2,L}^{n+1/2} = \tilde{\xi}_L -
  \underbrace{l_i^\evz \cdot \left [
        \tilde{\xi}_L  - \mathcal{I}^\evz_+(\xi_i)
       \right ] r_i^\evz}_{\text{only if~$u \ge 0$}}
\end{equation}
Again, ignoring flattening, if $u \ge 0$, then we have
\begin{equation}
\xi_{i+1/2,L}^{n+1/2} = \tilde{\xi}_L -
  \left (\tilde{\xi}_L  - \mathcal{I}^\evz_+(\xi_i) \right ) = \mathcal{I}^\evz_+(\xi_i)
\end{equation}
(where we used the fact that the eigenvectors are normalized to unity and
don't mix in any other states when dealing with passive terms).  This
is the expected behavior---we see a state that is traced only by the
contact wave.  If $u < 0$ but $u + c \ge 0$, then we instead get:
\begin{equation}
\xi_{i+1/2,L}^{n+1/2} = \tilde{\xi}_L = \mathcal{I}^\evp_+(\xi_i)
\end{equation}
Here we used the same definition of the reference state and see that our
interface state sees the profile traced under the fastest wave, not the
contact.  This is not the correct behavior for a passively-advected
quantity.  

The fix for passively-advected quantities is to simply ignore the 
idea of a reference state and just test on the speed of the contact
itself, setting:
\begin{equation}
\xi_{i+1/2,L}^{n+1/2} = \left \{ \begin{array}{cc}
       \mathcal{I}_+^\evz(\xi_i) & \mathrm{if~} u  > 0 \\
       \xi_i                    & \mathrm{otherwise}
\end{array}
\right .
\end{equation}

\subsection{Source Term Predictor for the Hydrodynamics}
\label{sec:source_term_predictor}

In the original release of \castro\ we used the time-level $n$ value of the 
gravitational and rotation source terms in constructing the edge-states for 
the hydrodynamics update. While this is formally second-order accurate, there
is a better choice one can make. We have information about the trend of these
source terms from previous timesteps, so we can use a predictor method to guess
at a more accurate value of the gravitational and rotational fields at the
$n+1/2$ time-level the hydro is evaluated at. Our method uses a lagged linear
extrapolation. Going into the hydro update, we have both the time-level $n$ and
time-level $n-1$ data for $g$ (as well as the acceleration due to rotation).
From this one can construct a simple linear estimator of (say) the gravitational
acceleration using a backward difference scheme at time-level $n$:
\begin{equation}
  \mathbf{g}^{n+1/2} \approx \mathbf{g}^n + \frac{\Delta t_n}{2} \frac{d\mathbf{g}^n}{dt} \approx \mathbf{g}^n + \frac{\Delta t_n}{2}\left(\frac{\mathbf{g}^n - \mathbf{g}^{n-1}}{\Delta t_{n-1}}\right).
\end{equation}

While both this method and the original method have second-order convergence properties,
we have seen from testing that this new method is slightly more accurate in absolute terms.
Finally, we note that this predictor is not applied in cases where we do not have a suitable
time-level $n-1$ value of the source term, such as in the first timestep of a simulation.

We note also that we have changed slightly how we use source terms in \castro's 
hydrodynamics. In the original release we would explicitly handle gravity, rotation, 
and user-defined external source terms separately in constructing the edge states. 
At present, we sum all of these source terms prior to starting the hydrodynamics 
update, and use a single source array with data for all components of the state. 
Consequently, we actually do the source-term predictor shown here on this source 
term array, rather than individually on each component of the forcing.

\subsection{Source Term Tracing}
\label{sec:gravity_rotation_tracing}

We note a few additional differences between the original PPM
implementation of \citet{ppm} and \castro.  In the original PPM
implementation, the gravitational acceleration was reconstructed as a parabola, and
this was traced under to find the forcing that affects the interface
for each wave.  \castro\ originally followed \citet{ppmunsplit} which instead adds
$(\Delta t/2)g$ to the interface states for velocity at the end of the
reconstruction.  In the current implementation, we return to the original
parabolic reconstruction and characteristic tracing. In fact, as described 
in \autoref{sec:source_term_predictor}, since we send to the hydro a single 
source term array that holds the sum of all the source terms (including gravity 
and rotation), we do the parabolic reconstruction on the full source term data. 
This can be controlled in \castro\ with the parameter {\tt castro.ppm\_trace\_sources}. 

For the following explanation of how the tracing works, we consider only gravity.
Our system with the source appears as:
\begin{equation}
q_t + A(q) q_x = G
\end{equation}
where $G = (0, g, 0)^T$---i.e. the gravitational source only affects
$u$, not $\rho$ or $p$.  Note that in the PPM paper, they put $G$ on 
the left-hand side of the primitive variable equation, so our signs are
opposite.  Our projections are now:
\begin{equation}
\sum_{\nu; \lambda^\enu \ge 0}l^\enu \cdot (\tilde{q} - \mathcal{I}^\enu_+(q) - \tfrac{\Delta t}{2} G) r^\enu
\end{equation}
for the left state, and
\begin{equation}
\sum_{\nu; \lambda^\enu \le 0} l^\enu \cdot (\tilde{q} - \mathcal{I}^\enu_-(q) - \tfrac{\Delta t}{2} G) r^\enu 
\end{equation}
for the right state.  Since $G$ is only non-zero for velocity, only
the velocity changes.  Writing out the sum (and performing the vector products), we
get:
\begin{eqnarray}
u_{i+1/2,L}^{n+1/2} =
   \tilde{u}_+ 
  &-& \frac{1}{2} \left [
      \left (\tilde{u}_+ - \mathcal{I}_+^\evm(u) - \frac{\Delta t}{2} \mathcal{I}^\evm_+(g) \right ) - 
       \frac{\tilde{p}_+ - \mathcal{I}_+^\evm(p)}{C} \right ] \nonumber \\
  &-& \frac{1}{2} \left [
      \left (\tilde{u}_+ - \mathcal{I}_+^\evp(u) - \frac{\Delta t}{2} \mathcal{I}^\evp_+(g) \right ) +
       \frac{\tilde{p}_+ - \mathcal{I}_+^\evp(p)}{C} \right ]
\end{eqnarray}
(The expression in the PPM paper contains $\Delta t G$, not $(\Delta t/2) G$,
but we believe that the factor of $1/2$ is correct.  To see this, notice that if both
waves are moving toward the interface, then the source term that is
added to the interface state is $(\Delta t/4) (\mathcal{I}_+^\evm(g) +
\mathcal{I}_+^\evp(g))$ for the left state, which reduces to $(\Delta
t/2) g$ for constant g---this matches the result from Taylor
expanding to the interface at the half-time (as in \citealt{ppmunsplit}).)

There is one additional effect of this change---now the gravitational
source is seen by all Riemann solves (including the transverse solves)
whereas previously it was only added to the final unsplit interface
states.  Both methods are second-order accurate.

\section{Proof of Energy Conservation in Simulations using Self-Gravity}
\label{app:gravity}

In \autoref{sec:gravity_hydro_coupling}, we described our approach to updating the gas energy
in response to motions of fluid through the self-generated gravitational potential using 
\autoref{eq:grav_energy_conservation_update}. While it is straightforward to observe that this approach
should be conservative for an arbitrary fixed external potential $\Phi$, it is not as obvious that this
should be so for a self-generated potential which changes in response to mass motions on the domain. To
see that this still holds for the self-generated gravitational potential $\Phi$, let us start with 
\autoref{eq:grav_energy_conservation_update} in a slightly revised form:
\begin{equation}
  \Delta(\rho E)_i = -\frac{1}{2}\sum_{j} \Delta\rho_{ij}(\Phi_i - \Phi_{j}) \label{eq:grav_energy_conservation_update_revised}
\end{equation}
where by $\Delta \rho_{ij}$ we mean the density transferred from zone $j$ to zone $i$, so that
$\Delta \rho_{ij} = - \Delta \rho_{ji}$, and the sum is over all zone indices $j$ that are adjacent
to zone $i$. Let us define $\Phi_{ij} = \Phi_{ji} = (\Phi_{i} + \Phi_{j}) / 2$ as the potential on the
zone interface between zones $i$ and $j$. Then we have:
\begin{equation}
  \Delta(\rho E)_i = -\sum_{j} \Delta\rho_{ij}(\Phi_i - \Phi_{ij}).
\end{equation}
We can evalute the sum for all of the terms proportional to $\Phi_i$ by observing that the change in
density from time-level $n$ to time-level $n+1$ is the sum of the density fluxes from all adjacent zones.
\begin{equation*}
  \Delta(\rho E)_i = - (\rho_i^{n+1} - \rho_i^{n}) \Phi_i + \sum_{j}\Delta \rho_{ij} \ \Phi_{ij}
\end{equation*}
Now let us sum this over all zones $i$ in the domain, and ignore the domain boundaries, or assume that they are
far enough away from the region of compact support for $\rho$ that $\Phi$ is negligible there. As the second
term on the right-hand side is antisymmetric in $i$ and $j$, it cancels when summing adjacent zones, and we have:
\begin{equation*}
  \sum_{i} (\rho E)_i^{n+1} - \sum_{i} (\rho E)_i^{n} = -\frac{1}{2}\sum_{i} (\Phi_{i}^{n+1} + \Phi_{i}^{n})(\rho_i^{n+1} - \rho_i^{n})
\end{equation*}
Note that, as explained the text, we are using a time-centered $\Phi$ to correspond to the mass fluxes
at time-level $n+1/2$. Finally we re-write this in a form where the difference in total energy between time-levels
$n$ and $n+1$ is on the left-hand side and any sources causing this to be non-zero are on the right-hand side:
\begin{align}
  \sum_{i} \left(\rho E + \frac{1}{2}\rho\Phi\right)_i^{n+1} - \sum_{i} \left(\rho E + \frac{1}{2}\rho\Phi\right)_i^{n} &= \frac{1}{2}\sum_{i} \left(\Phi_i^{n+1}\rho_i^{n} - \Phi_i^{n}\rho_i^{n+1}\right) \notag \\
       &= \frac{1}{8\pi G} \sum_{i}\left(\Phi_{i}^{n+1}\nabla^2 \Phi_{i}^{n} - \Phi_i^{n}\nabla^2 \Phi_{i}^{n+1}\right) \label{eq:total_energy_difference}
\end{align}
\autoref{eq:total_energy_difference} expresses total energy conversation if and only if the right-hand side vanishes.
We observe that the right-hand side has the form of a variant of the divergence theorem often called Green's second identity:
\begin{equation}
  \int (\Phi^{n}\nabla^2 \Phi^{n+1} - \Phi^{n+1}\nabla^2 \Phi^{n}) dV = \int \left(\Phi^{n} \nabla \Phi^{n+1} - \Phi^{n+1} \nabla \Phi^{n}\right) \cdot d\mathbf{S}, \label{eq:green_second_identity}
\end{equation}
where $d\mathbf{S}$ is the area element with vector component parallel to the outward normal. The analogous result holds for the
discretized form in \autoref{eq:total_energy_difference}. With the assumptions used above, the right-hand side of
\autoref{eq:green_second_identity} will vanish as the surface integral is evaluated at infinity, where the potential
tends to zero. This concludes the proof that the method is conservative when the potential used at the zone interfaces
is time-centered, even in light of the change of the potential over the timestep due to the mass motion that is causing the change in the energy.

From the above discussion it is straightforward to see exactly why the method is not fully conservative to machine
precision in practice. First, we cannot simulate the domain out to infinity, so Green's second identity does not hold exactly
and there is some loss or addition of energy at domain boundaries. Second, \autoref{eq:total_energy_difference} holds in the
continuum limit by using the Poisson equation, but in practice it is not exactly true that $\rho_i = 4\pi G \nabla^2 \Phi_{i}$ due
to small errors in the potential at the level of the tolerances used in the Poisson solver.

\section{Formulation of the Multipole Expansion for the Gravitational Potential}
\label{app:multipole}

The integral formulation of the gravitational potential, using a series expansion in spherical harmonics, is:
\begin{equation}
  \Phi(\mathbf{x}) = -G\sum_{l=0}^{\infty} \sum_{m=-l}^{l} \frac{4\pi}{2l+1} \int \rho(\mathbf{x}^\prime)\, Y_{lm}(\theta,\phi)\, Y_{lm}^*(\theta^\prime,\phi^\prime)\, \frac{r_{<}^{l}}{r_{>}^{l+1}}\, dV^\prime,
\end{equation}
where $\theta$ is the polar angle and $\phi$ is the azimuthal angle, $r \equiv |\mathbf{x}|$ is the radial distance, and at any point in the domain $r_{<}$ is the smaller of $r$ and $r^\prime$, and $r_{>}$ is the larger of the two. This immediately suggests writing the potential at any location as the sum of
two series:
\begin{equation*}
  \Phi(\mathbf{x}) = -G\sum_{l=0}^{\infty} \sum_{m=-l}^{l} \frac{4\pi}{2l+1}\left[ q^{L}_{lm}(\mathbf{x})\, r^{-l-1} + q^{U}_{lm}(\mathbf{x})\, r^{-l-1} \right] Y_{lm}(\theta,\phi),
\end{equation*}
where we have defined two multipole moments as integrals over the domain:
\begin{align}
  q^{L}_{lm}(\mathbf{x}) &= \int dV^\prime\, \rho(\mathbf{x}^\prime)\, Y^*(\theta^\prime,\phi^\prime)\, \Theta(r - r^\prime)\, {r^\prime}^{l} \\
  q^{U}_{lm}(\mathbf{x}) &= \int dV^\prime\, \rho(\mathbf{x}^\prime)\, Y^*(\theta^\prime,\phi^\prime)\, \Theta(r^\prime - r)\, {r^\prime}^{-l-1}.
\end{align}
$\Theta(r)$ is the standard step function, equal to one if the argument is positive and zero if the argument is negative. Geometrically,
$q^{L}(\mathbf{x})$ is an integral containing only mass interior to $|\mathbf{x}|$, and $q^{U}(\mathbf{x})$ is an integral containing only
mass exterior to $|\mathbf{x}|$. Provided that one has computed these two integrals for a point $\mathbf{x}$, one can use the series expansion
to calculate the potential at that point in principle to arbitrary accuracy by including higher order terms.
  
We prefer to work with solely real-valued quantities, and so we make use of the addition theorem for spherical harmonics \citep[Section 3.6]{jackson}:
\begin{align}
  \frac{4\pi}{2l+1} \sum_{m=-l}^{l} Y^*_{lm}(\theta^\prime,\phi^\prime)\, Y_{lm}(\theta, \phi) &= P_l(\text{cos}\, \theta) P_l(\text{cos}\, \theta^\prime) \notag \\
   &+ 2 \sum_{m=1}^{l} \frac{(l-m)!}{(l+m)!} P_{l}^{m}(\text{cos}\, \theta)\, P_{l}^{m}(\text{cos}\, \theta^\prime)\, \left[\text{cos}(m\phi)\, \text{cos}(m\phi^\prime) + \text{sin}(m\phi)\, \text{sin}(m\phi^\prime)\right].
 \end{align}
The $P_l(x)$ are the Legendre polynomials and the $P_l^m(x)$ are the associated Legendre polynomials. We construct them using a stable recurrence relation given known values for $l = 0$ and $l = 1$. We can then formulate the expansion in a different way:
\begin{align}
  \Phi(\mathbf{x}) &= -G\sum_{l=0}^{\infty} \left\{ Q_{l}^{(L,0)}(\mathbf{x})\, P_l(\text{cos}\, \theta)\, r^{-l-1} + Q_{l}^{(U,0)}(\mathbf{x})\, P_l(\text{cos}\, \theta)\, r^{l} \right. \notag \\
  &\hspace{0.7in} + \sum_{m=1}^{l} \left[ Q_{lm}^{(L,C)}(\mathbf{x})\, \text{cos}(m\phi) + Q_{lm}^{(L,S)}(\mathbf{x})\, \text{sin}(m\phi)\right] P_{l}^{m}(\text{cos}\, \theta)\, r^{-l-1} \notag \\
  &\hspace{0.7in} + \left. \sum_{m=1}^{l} \left[ Q_{lm}^{(U,C)}(\mathbf{x})\, \text{cos}(m\phi) + Q_{lm}^{(U,S)}(\mathbf{x})\, \text{sin}(m\phi)\right] P_{l}^{m}(\text{cos}\, \theta)\, r^{l} \right\}
\end{align}

The multipole moments now take the form:
\begin{align}
  Q_l^{(L,0)}(\mathbf{x}) &= \int P_l(\text{cos}\, \theta^\prime)\, \Theta(r - r^\prime)\, {r^{\prime}}^l \rho(\mathbf{x}^\prime)\, d^3 x^\prime \\
  Q_l^{(U,0)}(\mathbf{x}) &= \int P_l(\text{cos}\, \theta^\prime)\, \Theta(r^\prime - r)\, {r^{\prime}}^l \rho(\mathbf{x}^\prime)\, d^3 x^\prime \\  
  Q_{lm}^{(L,C)} &= 2\frac{(l-m)!}{(l+m)!} \int P_{l}^{m}(\text{cos}\, \theta^\prime)\, \text{cos}(m\phi^\prime)\, \Theta(r - r^\prime)\, {r^\prime}^l \rho(\mathbf{x}^\prime)\, d^3 x^\prime \\
  Q_{lm}^{(U,C)} &= 2\frac{(l-m)!}{(l+m)!} \int P_{l}^{m}(\text{cos}\, \theta^\prime)\, \text{cos}(m\phi^\prime)\, \Theta(r^\prime - r)\, {r^\prime}^{-l-1} \rho(\mathbf{x}^\prime)\, d^3 x^\prime \\
  Q_{lm}^{(L,S)} &= 2\frac{(l-m)!}{(l+m)!} \int P_{l}^{m}(\text{cos}\, \theta^\prime)\, \text{sin}(m\phi^\prime)\, \Theta(r - r^\prime)\, {r^\prime}^l \rho(\mathbf{x}^\prime)\, d^3 x^\prime \\
  Q_{lm}^{(U,S)} &= 2\frac{(l-m)!}{(l+m)!} \int P_{l}^{m}(\text{cos}\, \theta^\prime)\, \text{sin}(m\phi^\prime)\, \Theta(r^\prime - r)\, {r^\prime}^{-l-1} \rho(\mathbf{x}^\prime)\, d^3 x^\prime.  
\end{align}
In practice, of course, we select some maximum value $l_{\text{max}}$ at which we terminate the summation, determined either by computational efficiency requirements or by the fact that there is little information at high orders for sufficiently smooth mass distributions. In \castro\ we have the capability to compute any of the above multipole moments, though in this paper we are only using the multipole expansion to calculate the boundary conditions on the potential, and so we neglect calculation of the moments with a $U$ subscript as we are assuming that all of the mass is interior to the boundary. \autoref{eq:multipole_potential} is directly recovered under these conditions.

\clearpage

\bibliographystyle{aasjournal}

\clearpage

\end{document}